\documentclass[twocolumn,aps,tightenlines,floats,floatfix]{revtex4}

\def\bea{\begin{eqnarray}}
\def\eea{\end{eqnarray}}
\def\etal{{\it et.\ al.\/},$\,$}

\newcommand{\bra}[1]{\left\langle #1\right|}
\newcommand{\ket}[1]{\left|#1\right\rangle}

\newcommand{\gammavec}{\mbox{\boldmath$\gamma$}\,}

\def\be{\begin{equation}}
\def\ee{\end{equation}}
\def\ba{\begin{eqnarray}}
\def\ea{\end{eqnarray}}

\usepackage{graphics}
\usepackage{graphicx}
\usepackage{epsf} 
\usepackage{amsmath}
\usepackage{amssymb}
\usepackage{bbold}
\def\sfrac#1#2{{\textstyle \frac{#1}{#2}}}

\begin{document} 


\phantom{0}
\vspace{-0.2in}
\hspace{5.5in}
\parbox{1.5in}{ \leftline{JLAB-THY-06-500}} 

\vspace{-1in}
\title
{\bf A pure $S$-wave covariant model for the nucleon}
\author{Franz Gross$^{1,2}$, G. Ramalho$^{2,3}$ and M. T. Pe\~na$^{3,4}$
\vspace{-0.1in}  }

\affiliation{
$^1$College of William and Mary, Williamsburg, Virginia 23185 \vspace{-0.15in}}
\affiliation{
$^2$Thomas Jefferson National Accelerator Facility, Newport News, 
Virginia 23606 \vspace{-0.15in}}
\affiliation{
$^3$Centro de F{\'\i}sica Te{\'o}rica de Part{\'\i}culas, 
Av.\ Rovisco Pais, 1049-001 Lisboa, Portugal \vspace{-0.15in}}
\affiliation{
$^4$Department of Physics, Instituto Superior T\'ecnico, 
Av.\ Rovisco Pais, 1049-001 Lisboa, Portugal}

\date{\today}
\begin{abstract} 

Using the manifestly covariant spectator theory, and modeling the nucleon as a system of three constituent quarks with their own electromagnetic structure,  we show that all four nucleon electromagnetic form factors can be very well described by a manifestly covariant nucleon wave function with {\it zero\/} orbital angular momentum.  Since the concept of wave function depends on the formalism,  the conclusions of light-cone theory requiring nonzero angular momentum components are not inconsistent with our results.   We also show that our model gives a qualitatively correct description of deep inelastic scattering, unifying the phenomenology at high and low momentum transfer.  Finally we review two different definitions of nuclear shape and show that the nucleon is spherical in this model, regardless of how shape is defined.   


\end{abstract}
\phantom{0}
\vspace*{0.9in}  
\maketitle


\section{Introduction}

The elastic electron-proton polarization transfer experiments undertaken at the Thomas Jefferson National Accelerator Facility
(JLab) \cite{Jones99,Gayou01,Punjabi05}  disclosed that the ratio of the
electric $G_{Ep}$ to magnetic $G_{Mp}$ form factors of the proton is not constant as $Q^2$, the square of the
momentum transfer, varies (referred to as a lack of scaling).   These were followed up by the super Rosenbluth measurements at JLab \cite{Christy:2004rc} confirming  older Stanford Linear Accelerator Center (SLAC) measurements (reanalyzed by Arrington \cite{Arrington:2003df}) which originally showed that $G_{Ep}$ and $G_{Mp}$ {\it do\/} scale.

It now seems clear that the 
discrepancy between the 
JLab polarization transfer results  and other measurements 
using Rosenbluth separation are essentially due to 
two-photon processes \cite{Arrington07,Perdrisat06,Arrington06,Carlson07}.
There are still some uncertainties about 
the effective sizes of two-photon exchange effects but 
there is no doubt that 
the form factors extracted from the Rosenbluth method require 
significant corrections, while those extracted from the polarization transfer method require 
only small corrections. 
As a consequence the polarization transfer method is a more accurate way to determine 
the form factors, 
removing the possibility that the lack 
of scaling is an experimental artifact.

The lack of scaling was a surprise.  While it had been predicted as long ago as 1972 \cite{Iachello:1972nu} these predictions had been largely forgotten.   The absence of scaling was seen by some as proof that the quark wave function of the proton {\it  must\/} have orbital angular momentum components $L>0$, and this observation was further supported by arguments, based on light-cone wave functions, that the Pauli form factor, $F_2$, {\it must\/} be zero unless $L>0$ components exist \cite{Brodsky:1997de}.   In this paper we will discuss how these results depend on the light-cone formalism and show, using the covariant spectator formalism, that  (i) it is possible to construct a pure $S$-wave covariant wave function for the nucleon, and that (ii) $L>0$ components, while they may be a part of any realistic wave function, are  {\it not required\/} to explain the data.  We cannot conclude that either formalism is wrong, only that the  the concept of the ``wave function'' is different.  All of this is discussed in Sec.~\ref{sec:j}.

The JLab data has also stimulated discussion 
about the shape of the nucleon.  Is the 
nucleon spherical or it is deformed?
If distorted, is this a relativistic effect?  The answers to these questions depend in part on how we define ``shape'' and will be discussed in some detail in Sec.~\ref{sec:def}.

Finally, in order to fully examine the implications of our covariant approach,  we use it to calculate the quark distribution functions in deep inelastic scattering (DIS).  We obtain reasonable, qualitatively correct, results.

\begin{figure*}
\centerline{
\mbox{
\includegraphics[width=6in]{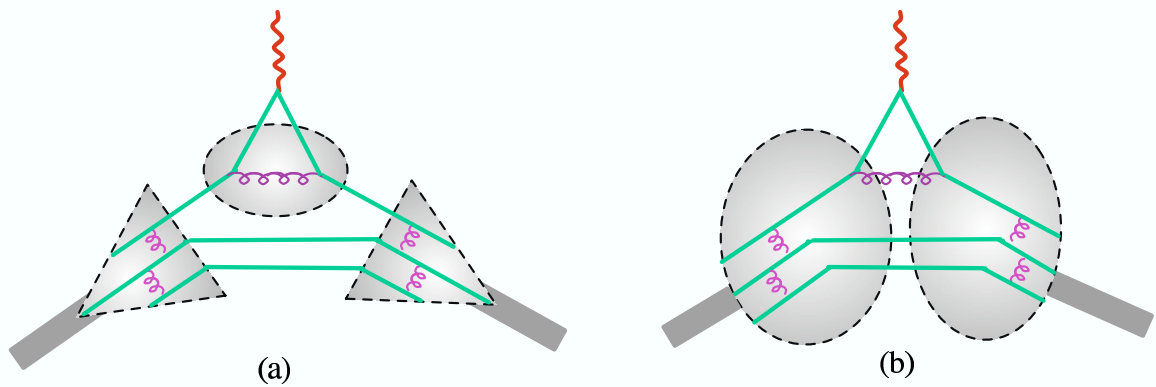} }}
\caption{\footnotesize{(Color on line) Two views of the nucleon form factor.  Left panel (a): the form factor is built from a constituent quark form factor and nuclear wave functions.  Right panel (b): the quark is point-like, with its ``structure'' described by higher Fock states of the nuclear wave function; in this case the Fock state with 3-quarks and one gluon. }}
\label{fig:one}
\end{figure*}

To set the stage for the physical approximations used in this calculation, consider Fig.~\ref{fig:one}.  In the $N_c\to\infty$ limit (where $N_c$ is the number of quark colors), crossed diagrams are suppressed and the gluon exchanges between the $q\bar q$ pair interacting with the photon can be included either as part of a constituent quark electromagnetic form factor (usually described by vector dominance)  or as contributions from higher Fock components of the nuclear wave function.   (Since $N_c=3$, corrections to this simple picture are expected to be of the order of only about 10\%, and are probably small enough to be included when the parameters of our  phenomenological model are adjusted to fit the data.)  At low energies and momenta, the description of Fig.\ \ref{fig:one}(a) has the advantage that unknown short-range physics can be included in a few constituent quark parameters (such as the anomalous moment).   In this paper we will adopt the view represented by Fig.\ \ref{fig:one}(a); our constituent quark form factors include the physics that comes from higher Fock components, expected to be important at modest $Q^2$, vanishing only in the DIS limit.

The remainder of this paper is divided into five sections. 
In the next section we define the nucleon wave function in some detail.  
In Sec.~III, we derive the nucleon form factors using this wave function.  The principal results are Eqs.~(\ref{eqF1}) and (\ref{eqF2}).  Readers uninterested in the details may skip directly to Sec.~IV, where the first qualitative results that can be derived from these equations are discussed.  Then, before studying the form factors quantitatively, we use this model to derive, in Sec.~V, formulas for the quark distribution amplitudes measured in deep inelastic scattering.  In Sec.~VI we parameterize both the nucleon wave function and the quark form factors and determine the parameters by fitting the nucleon form factors and the moments of the quark distribution amplitudes.  Finally, in Sec.~VII, we review the work and draw conclusions.  The appendices provide a few details considered too technical to be included in the text.

\section{Relativistic Nucleon wave function}\label{sec:wf}

The relativistic nucleon wave function presented here is similar to the one developed by Gross and Agbakpe \cite{Gross06}, referred to here as Ref.~I.  Two observations motivated us to develop this new $S$-wave model.  First, the overall motion of the composite nucleon described in Ref.~I does not satisfy the Dirac equation for a spin-1/2 particle (we are not sure that this is an essential requirement, but it is certainly an advantage). Attempts to find a wave function that does satisfy the Dirac equation lead us directly to this model.   Second, as emphasized by Kvinikhidze and Miller~\cite{Miller06}, the model of Ref.~I depends on the {\it direction\/} of the relative three-momentum ${\bf k}$ between a pair of {\it noninteracting\/} quarks (referred to as a diquark although this term is usually applied to an {\it interacting\/} pair) and the third quark, and hence includes angular momentum components.  The $S$-wave model presented here is {\it independent of the direction of this momentum\/}, and {\it exactly\/} reproduces the pure $S$-wave structure of the simplest  nonrelativistic $SU(2)\times SU(2)$ wave function when the nucleon is at rest.  Since one of the goals of this paper is to study the implications of the absence of $L>0$ components in the wave function, it is essential that our model wave function have no such components.
 
Following Ref.~I \cite{Gross06},
we consider the nucleon described by a wave function built
in the spirit
of the covariant spectator formalism \cite{Gross69,Gross82},
which has already been solved and shown to work successfully for the three-nucleon system \cite{Gross82,Stadler97a,Stadler97b}. 
The nucleon with four-momentum $P$ and mass $M$  is described by a wave function for an off-shell quark  
and an on-mass-shell diquark like cluster. 
\be
\Psi(P,k)=(m_q-\not\! p_1)^{-1} 
\bra{k} \Gamma  \ket{P}\label{Nwf0}
\ee
where $\Gamma$ is the vertex function describing the coupling of an incoming on-shell nucleon with mass $M$ to an outgoing off-shell quark and an on-shell quark pair (the ``diquark'').   The continuous mass of the diquark pair is fixed at some mean value, which scales out of the final results. The quark has dressed mass $m_q$ and four-momentum $p_1$.
The diquark four-momentum $k=P-p_1$
is constrained by its on-mass-shell condition $k^2=m_s^2$,
where $m_s$ is the mass of the diquark. 

As discussed in Ref.~I, the wave function will be parameterized by a smooth function with no singularity at the quark pole $p_1^2=m_q^2$.  If the mass of the quark and diquark were greater than the mass $M$ of the nucleon, there would be no pole in any case, but if $m_q+m_s<M$ the propagator would normally contribute a pole associated with the possibility of free scattering of the quark and diquark.  Since the quarks are confined, this scattering cannot take place, and detailed calculations \cite{Savkli} show that the  vertex function $\Gamma$ automatically develops a zero that cancels the quark pole and gives a smooth behavior for the wave function.  Our assumption that $\Psi$ is smooth regardless of the value of the quark mass is a simple way of including confinement without investigating the details of this cancellation.

The quark-diquark interaction described by the wave function is parametrized 
through simple scalar wave functions. 
The  diquark can have either spin-0 (isospin-0) and 
spin-1 (isospin-1) components. 
The isospin states of the quark-diquark system can be written
\ba
 \phi_I^0&=&\xi^{0*} \chi^{I} \label{iso0}\\
 \phi^1_{I}&=&-\sfrac{1}{\sqrt{3}} \tau \cdot \xi^{1*} \chi^{I}\nonumber\\
 &=&\sfrac{1}{\sqrt{6}}\left[ \tau_-\xi^1_+ -\tau_+\xi^1_ - - \sqrt{2}\,\tau_3\xi^1_0\right]\chi^{I}
\label{iso1}
\ea
where $\tau_\pm=\tau_x\pm i\tau_y$ are the isospin raising and lowering operators, $I=\pm1/2$ is the isospin of the quark (or nucleon)
\ba
\chi^{+\frac12}=\left(\begin{array}{c} 1\cr 0\end{array}\right)=u\,({\rm or}\,p) \quad
\chi^{-\frac12}=\left(\begin{array}{c} 0\cr 1\end{array}\right)=d\, ({\rm or}\,n) , \quad\label{quark}
\ea
 and  
\ba
& & \xi^0=\sfrac{1}{\sqrt{2}} (ud-du)  \nonumber \\
& &\xi^{1}_0 =\sfrac{1}{\sqrt{2}} (ud+du) =
 \xi_z \nonumber \\
& & \xi^{1}_+= uu =-\sfrac{1}{\sqrt{2}}(\xi_x+i\xi_y)\nonumber\\
&& \xi^{1}_-=dd =\sfrac{1}{\sqrt{2}}(\xi_x-i\xi_y)\; .
\label{diquark}
\ea
The operator $\tau\cdot\xi^{1*}$ in Eq.\ (\ref{iso1}) is to be interpreted as transforming the initial two-component nucleon spinor into a two-component  quark spinor.  Explicitly, for the proton state this operation gives
\ba
\phi^1_{\frac{1}{2}}&=&\sqrt{\sfrac23}
\left(\begin{array}{c} 
0\cr 
1\end{array}\right)
\xi^1_+ -\sqrt{\sfrac13}
\left(\begin{array}{c} 
1\cr 
0\end{array}\right)
\xi^1_0\nonumber\\
&\to& \sqrt{\sfrac16}\left[2d(uu) -
u(ud+du)\right],
\ea
and automatically yields the correct quark content of the isospin-1 diquark part of the proton wave function.

In the nucleon rest frame, we choose spin states that are analogs of (\ref{iso0}) and (\ref{iso1}).  This means that the spins of both the quark and diquark are aligned along the (arbitrarily chosen) $z$-axis.  [In Ref.~I the spin of the diquark was chosen to be parallel to the diquark momentum, ${\bf k}$, introducing an angular dependence into the wave function from the start.  Here we construct a wave function with {\it no\/} angular dependence by defining the spin with respect to a fixed axis.]   For a nucleon moving in an arbitrary direction, the spins will be aligned along the direction of the nucleon momentum.  This means that the spin states must be constructed in a two-step process: (i) first the nucleon is transformed to momentum $|{\bf P}|=\mbox{P}$ along the $\hat z$ direction, and then (ii) the  state is transformed back to the final direction ${\bf P}$.  This construction bears some resemblance to the construction of helicity amplitudes.  A more complete discussion of how these states are constructed can be found in  Refs.~\cite{FApaper,RPG}.  

In this paper we will need explicit results  for nucleons moving in the $\pm\hat z$ direction only.  The scalar diquark state can be written  
\be
\phi_s^0
= u({\bf 0},s) \to u({\bf P},s),
\label{eqphi0}
\ee
where $\phi^0_{\pm 1/2}$ is the spin-0 diquark state, $u$ the Dirac spinor, 
and the arrow indicates the relativistic generalization.
Note that, in this case, the spin of the (isolated) quark  is specified by the spin of the nucleon.

To construct the spin-1 diquark component of the wave function, $\phi_s^1$, we begin by considering the nucleon at rest, where the lower two components of its Dirac spinor are zero.  If we choose 
\be
\phi_s^1= -\frac{1}{\sqrt{3}}\gamma_5\not\!\varepsilon^*\,
u({\bf 0},s)
\label{phi1}
\ee
with the four components $\varepsilon=\{\varepsilon_t,\varepsilon_x,\varepsilon_y,\varepsilon_z\}$ of the three diquark polarization states defined as in Eq.\ (\ref{diquark}) 
\ba
\varepsilon_\pm&=&\mp\sqrt{\sfrac{1}{2}}\left\{0,1,\pm i,0\right\} \nonumber\\
\varepsilon_0&=&\left\{0,0,0,1\right\},
\ea
then the upper two components of (\ref{phi1})  will look just like (\ref{iso1}) and the lower two components will be zero.    The state in the moving frame is then obtained from (\ref{phi1}) by a boost, and becomes
\ba
\phi_s^1\to -\frac{1}{\sqrt{3}}\gamma_5\not\!\varepsilon_P^*\,
u({\bf P},s) \label{phi1R}
\ea
where, for boosts in the $\hat z$ direction, the $\varepsilon_\pm$ polarization vectors are unchanged, but the longitudinal one becomes
\ba
\varepsilon_{0 P}&=&\frac{1}{M}\left\{\mbox{P},0,0,E_P\right\},
\ea
where $E_P=\sqrt{M^2+\mbox{P}^2}$.  Note that all the polarization vectors satisfy
\ba
\varepsilon_\lambda^*\cdot \varepsilon_{\lambda'}=-\delta_{\lambda\lambda'}\qquad P\cdot\varepsilon_\lambda=0\,
\ea
(where we use the metric $g_{00}=1$ and $g_{ij}=-\delta_{ij}$).  The latter condition is the usual constraint ensuring the polarization vectors have only three independent states.  Finally, we note that we chose to write (\ref{phi1R}) in terms of $\varepsilon^*$ instead of  $\varepsilon$ to allow us to interpret (\ref{phi1R}) as amplitude for an incoming nucleon and an {\it outgoing\/}  diquark in the final state.

Putting this all together, the manifestly covariant nucleon wave function is the four-component Dirac spinor 
\ba
\Psi_N (P,k)&=&
\frac{1}{\sqrt{2}} \psi_0(P,k) \, \phi^0_I \, u({\bf P},s) 
\label{psiRel} \nonumber\\
& &-\frac{1}{\sqrt{6}} \psi_1(P,k) \; 
\phi^1_I \, 
 \gamma_5 \not\!\varepsilon^*_P \,u({\bf P},s),  \qquad \label{Nwf00}
\ea
which is a sum of contributions from a spin-isospin (0,0) diquark and a spin-isospin (1,1) diquark and 
$\psi_{0,1}$ are scalar functions that specify 
the relative shape of the two components.  
If $\psi_0=\psi_1=\psi$, Eq.\ (\ref{psiRel}) reduces to 
\be
\Psi_{N}=
\sfrac{1}{\sqrt{2}}(\phi_I^0 \phi_s^0+ \phi_I^1 \phi_s^1) \psi,
\label{psiNR}
\ee
in precise agreement with the symmetric nonrelativistic $SU(2)\times SU(2)$ wave function of the nucleon.  We emphasize that the combination $(\phi_I^0 \phi_s^0+ \phi_I^1 \phi_s^1)$ is exactly symmetric under interchange of any two quarks.  Note that, because $\not\!P$ commutes with $ \gamma_5\!\!\! \not\!\varepsilon^*$,  $\Psi_N$ satisfies the Dirac equation
\ba
(M-\not\!P)\Psi_N&=&\sfrac{1}{\sqrt{2}} \Big(\psi_0\phi_I^0 -
\frac{1}{\sqrt{3}}  \psi_1\phi^1_I 
 \gamma_5\!\! \not\!\varepsilon^*\Big)
 \nonumber\\
 &&\quad\times(M-\not\!P) \,u({\bf P},s) = 0. \label{Dirac}
\ea

The wave functions $\psi$ are Lorentz scalars that, by the Hall-Wightman theorem, can only depend on scalar products of their arguments, and since $k^2=m_s^2$ and $P^2=M^2$ are fixed, they can therefore only be a function of $(P-k)^2$.  We chose to 
express this $(P-k)^2$ dependence in terms 
of the dimensionless variable
\be
\chi=\frac{(M-m_s)^2-(P-k)^2}{M m_s}\, ,\label{chi}
\ee
and take a functional form for $\psi$ that 
reduces to the Hulthen form (difference of two Yukawa functions) in the nonrelativistic limit,
and has an asymptotic $1/Q^2$ dependence for large $Q^2$, as expected 
from pQCD calculations  of the electromagnetic
form factors \cite{Carlson86,Carlson98b}.  This form is
\be
\psi(P,k)=\frac{N_0}{m_s(\beta_1+\chi)(\beta_2+\chi)}\, ,
\label{eqpsi0}
\ee
where $\beta_1$, $\beta_2$ are range 
parameters (we assume $\beta_2 > \beta_1$)
and $N_0$ is a normalization constant. 

We emphasize that, in the nucleon rest frame, the wave function (\ref{psiRel}) contains absolutely {\it no\/} angular dependence of any kind.   In Ref.\ I,  the diquark polarization vectors, which were denoted by $\eta$, depended on the diquark momentum $k$ with the property $\eta\cdot k=0$.

\section{Nucleon Electromagnetic Form Factors}

The calculation of the form factors, based on Fig.~\ref{fig:one}(a), parallels the discussion of Ref.~I.   The relativistic impulse approximation (RIA) to the nucleon current in the covariant spectator theory is \cite{Gross69,Gross82,Stadler97a,Stadler97b,Gross04b,Adam98,Gross:1987bu}
\ba
J_I^\mu 
& =&
3 \sum_{\varepsilon}
\int_k \bar \Psi_N (P_+,k) \;
j_I^\mu \Psi_N(P_-,k) 
\label{eqCurrent1} \\
&=&
\bar u(P_+) \left[ F_1(Q^2) \gamma^\mu 
+ F_2(Q^2) \frac{i \sigma^{\mu \nu} q_\nu}{2M} \right] u(P_-),
\nonumber 
\ea
where the spectator formalism places the diquark on its mass-shell.  Our ``diquark'' is  actually  two noninteracting quarks with a continuous mass distribution from $2m_q$ to $\infty$, so what we are doing here is averaging the integral over this mass distribution by fixing the mass at a {\it mean value\/} equal to $m_s$, which becomes a parameter of the model.  (We can do this because none of the physics depends strongly on the details of this mass distribution.)   With this restriction the four-dimensional loop integral reduces to an integration over the three-momentum of the on-shell spectator diquark
\be
\int_k= \int 
\frac{d^3 k}{(2\pi)^3 2 E_s}
\label{intk}
\ee
with $E_s=\sqrt{m_s^2+{\bf k}^2}$ the energy of the on-shell diquark,
and the sum is over the polarizations $\varepsilon$ of the spin-1 diquarks (see below).   [The scalar diquark term has no sum.]  The RIA neglects any exchange current contributions that might be present.  As a consequence of our definitions (\ref{chi}) and (\ref{eqpsi0}), the momentum $k$ may be scaled by the diquark mass $m_s$, giving final results independent of $m_s$.
The factor of 3 sums up the contributions from the three quarks.  The expression is covariant and may be written in any frame, but is most conveniently evaluated in the Breit frame with the initial ($P_-$) and final ($P_+$) four-momentum of the nucleons chosen to be
\ba
& &P_+=(E,0,0,Q/2) \nonumber \\
& &P_-=(E,0,0,-Q/2) \nonumber \\
& &q=(0,0,0,Q), \nonumber
\ea
with $Q=\sqrt{-q^2}$ the transferred four-momentum and $E=\sqrt{M^2+Q^2/4}$.   The spin indices of the  nucleons have been suppressed.

The electromagnetic coupling of a spin $1/2$ quark 
with a photon is written
\be
j_I^\mu=j_1 \left(\gamma^\mu -\frac{\not\!q\,q^\mu}{q^2}\right) +j_2  \frac{i \sigma^{\mu \nu}q_\nu}{2M}\, ,
\label{eqmiccur}
\ee
where the subtraction term proportional to $\not\! q  \, q^\mu$ is zero for the elastic form factors, but ensures that the current is automatically conserved in the deep inelastic limit.  The use of this term for DIS was justified in Ref.~\cite{BG1}, and will be discussed in more detail in future work.
The functions $j_1$ and $j_2$ are operators in the quark isospin space.  For $i=1,2$,
\be
j_i=\sfrac16 f_{i+}(Q^2)+\sfrac12f_{i-}(Q^2)\tau_3 
\label{quarkff}
\ee
where $f_{i\pm}$ are the isoscalar and isovector combinations, related to the $u$ and $d$ quark form factors by
\bea
\sfrac23 f_{iu}&=& \sfrac16 f_{i+} + \sfrac12 f_{i-} \cr
-\sfrac13 f_{id}&=& \sfrac16 f_{i+} - \sfrac12 f_{i-} \, .
\label{quarkff2}
\eea
The form factors are normalized (with $a=\{u,d\}$) to
\bea
f_{1a}(0)=1&\qquad& f_{2a}(0)=\kappa_a  \nonumber\\
f_{1\pm}(0)=1&\;\;\;\;& f_{2\pm}(0)=\kappa_\pm
\eea
where $\kappa_u$ and $\kappa_d$ are the $u$ and $d$ quark anomalous magnetic 
moments (scaled by the quark charges) and 
\bea
\kappa_+&=&  2\kappa_u - \kappa_d \nonumber\\
\kappa_-&=& \sfrac23 \kappa_u+\sfrac13 \kappa_d \, .\eea
Note that $\kappa_+=\kappa_-$ implies that $\kappa_u=\kappa_d$.
In all models presented here, we construct $f_{1\pm}$ and $f_{2\pm}$ to satisfy the conditions
\bea
\lim_{Q^2\to \infty} f_{1\pm}(Q^2)& \to & \lambda >1\nonumber\\
\lim_{Q^2\to \infty} f_{2\pm}(Q^2)& \to &0\, .
\label{asy}
\eea
The role of $\lambda$ will be discussed later. 
The quark form factors (\ref{quarkff2}) parameterize the charge and magnetic structure of the $u$ and $d$ CQ. 

The integral (\ref{eqCurrent1}) is evaluated by substituting the nucleon states (\ref{psiRel}) and summing over diquark polarizations.  Since the diquark is a free particle, its polarization state cannot be changed by the interaction with the quark, so there is no coupling between the scalar and vector diquarks, giving an expression of the form 
\ba
& &J_I^\mu
=\bar u(P_+) 
\sfrac{3}{2} \int_k \left\{
\left[j_1 \gamma^\mu +j_2  \frac{i \sigma^{\mu \nu}q_\nu}{2M}\right] \psi_0^+ \psi_0^-  
\begin{matrix} \cr \cr \end{matrix} 
\right. \label{current} \\
& & \left.
-\sfrac{1}{3} 
\gamma_\nu \gamma_5 
\left[j_3 \gamma^\mu +j_4 \frac{i\sigma^{\mu \alpha} q_\alpha}{2M}
\right] \gamma_5 \gamma_{\nu^\prime} 
D^{\nu \nu^\prime} \psi_1^+ \psi_1^-
\right\} u(P_-),
\nonumber 
\ea
where $\psi_{0,1}^\pm= \psi_{0,1}(P_\pm,k)$ (and we allow for the possibility that $\psi_0\ne\psi_1$ for the time being) and the isospin sum has been done, giving $j_{(i+2)} \equiv \frac{1}{3} \tau_j j_i \tau_j= 
\sfrac16 f_{i+}- \frac{1}{6} \tau_3 f_{i-}$, for $i=1,2$.   The operator $D^{\nu\nu'}$ results from the sum of the vector polarizations of the diquarks.   
 
The derivation of the operator $D^{\nu \nu'}$ requires careful discussion.  In the first version of this paper \cite{Gross:2006fg} our derivation of this polarization sum was carried out in the Breit frame and Kvinikhidze and Miller~\cite{Kvinikhidze:2007qq} claimed it was not covariant.  
A new, more complete and careful derivation of  
$D^{\mu\nu}$ is given in Refs.~\cite{FApaper,RPG}, 
where we show in detail that $D^{\mu\nu}$ is both covariant and uniquely 
defined.    The derivation involves the introduction 
of new diquark (and quark) polarization states which 
can be described as ``covariant fixed axis polarization states.''   Discussion of these states requires the kind of care used in the original derivation of helicity states given by Jacob and Wick \cite{Jacob:1959at}.   A general discussion of the use of covariant fixed axis polarization states is planned for future work.  

Using these states, the covariant polarization sum $D^{\mu\nu}$ is
\ba
D^{\mu \nu}
& \equiv &
\sum_\lambda \varepsilon_{\lambda P_+}^\mu
\varepsilon_{\lambda P_-}^{\nu*}
\nonumber\\
\quad &=& -g^{\mu \nu} -\frac{P_+^\mu P_-^\nu}{M^2}
+2 \frac{(P_++P_-)^\mu (P_++P_-)^\nu}{4 M^2+Q^2}.\qquad
\label{eqDeltaP2}
\ea
Note that this function $D^{\mu\nu}$ has {\it no\/}  angular dependence, and the nuclear current is {\it pure\/} $S$-wave when $Q^2=0$.  At nonzero $Q^2$ an angular dependence emerges from the wave functions $\psi^\pm$, but this is due to the distortion under the boost and is not associated with the intrinsic structure of the state.

The two form factors $F_1$ and $F_2$ can be separated from the expression (\ref{current})  using (\ref{eqDeltaP2}).  We now impose the condition $\psi_0=\psi_1=\psi$ as required by the symmetry of the state (this simple condition replaces Eq.\ (10) needed in Ref.\  I).    Instead of reporting $F_1$ and $F_2$, we give the charge and magnetic combinations:
\bea
&&\!\!\!G_E(Q^2)=F_1(Q^2)-\tau F_2(Q^2)\nonumber\\
&&=\sfrac12 B(Q^2)\Big\{\left(f_{1+}+\tau_3f_{1-}\right)-\tau \left(f_{2+}+\tau_3f_{2-}\right)\Big\}
\label{eqF1}\\ 
&&\!\!\!G_M(Q^2)=F_1(Q^2)+F_2(Q^2)\nonumber\\
&& =\sfrac16 B(Q^2)\Big\{\left(f_{1+}+5\tau_3f_{1-}\right) + \left(f_{2+}+5\tau_3f_{2-}\right)\Big\}\, , \quad\quad\label{eqF2}
\eea
where $\tau\equiv Q^2/(4M^2)$ is not to be confused with $\tau_3$.  
In these expressions the $Q^2$ dependence of the quark form factors has been suppressed, and we have introduced the covariant nucleon body form factor
\be
B(Q^2)=\int_k \psi^+\psi^-\label{body}
\ee
The simple factorization into a product of a body form factor times combinations of quark form factors is possible because only the 
scalar wave functions depend on 
the integration variable; the rest of the integrand
(the $j_i$ functions and their coefficients) 
are functions of $Q^2$ only.   We emphasize that this calculation of the form factors is manifestly covariant at every step.  

At $Q^2=0$ the charge form factor reduces to 
\be
G_E(0) = \frac{1}{2} \left[1 + \tau_3 \right] B(0)   = 
\frac{1}{2} \left[1 + \tau_3 \right]\, , 
\ee
if we impose the normalization condition
\be
B(0)=\int_k \psi^2 =1\, .\label{norm}
\ee
This equality fixes $N_0$.  Our theory is not complete enough to automatically fix the normalization (as in Refs.\  \cite{Adam98,Gross04b}), but we do not regard $N_0$ as an additional parameter.   
Note that the conditions (\ref{asy}) ensure that both $G_E$ and $G_M$ 
have the correct  asymptotic behavior ( $\sim 1/Q^4$ times logarithm 
corrections), provided 
$B(Q^2)\sim 1/Q^4$ (times logarithm corrections), and this 
is easily achieved through a judicious choice of the 
high momentum dependence  
of the scalar wave function $\psi$.

\section{First predictions for the form factors}

Some of the predictions of this model are easily obtained from the formulas (\ref{eqF1}) and (\ref{eqF2}) without fitting or detailed analysis.  

\subsection{Magnetic moments}
\
The magnetic moments are  given by
\bea
\mu_p&=& 1 + \sfrac16 (\kappa_+  + 5 \kappa_-)\nonumber\\
\mu_n&=& -\sfrac23 + \sfrac16(\kappa_+  - 5 \kappa_-)
\eea 
The simplest assumption that $\kappa_+=\kappa_-=\kappa$ gives immediately the well known quark model relation
\be
\frac{\mu_p}{\mu_n}=-\frac{3}{2}\, ,
\ee
but the proton moment (for example) will not be even approximately correct unless $\kappa\sim2$
\be
\mu_p=1+\kappa\simeq 3\, .
\ee
This implies that the $u$ and $d$ anomalous moments are also equal to 2.  A physical argument for this result will be given in Sec.~\ref{sec:j} below.

In this work we choose to reproduce the nucleon magnetic moments exactly, giving 
\bea
\kappa_+&=&3(\mu_p+\mu_n)-1=1.639\nonumber\\
\kappa_-&=&\sfrac35 (\mu_p-\mu_n)-1 =1.823\, ,
\label{isomoments}
\eea
or
\bea
\kappa_u= 1.778 \qquad \kappa_d=1.915
\label{quarkmoments}
\eea

\subsection{Neutron form factor} \label{sec:nff}

Equation~(\ref{eqF1}) shows that $G_{En}$ is identically zero if $f_{1+}=f_{1-}$ and $f_{2+}=f_{2-}$ (a consequence of vector dominance if they could be approximated by a single pole at either the $\rho$ or $\omega$ mass).  Since $\kappa_+\ne\kappa_-$, $G_{En}$ is not identically zero, but is very small.  

There are three possible ways to increase $G_{En}$.  One possibility is to add angular momentum components to the nucleon.  This will be discussed in a subsequent paper.  Two remaining possibilities are (i) to break the $f_{1+}=f_{1-}$ equality, or (ii) to add a pion cloud term.  These will be discussed in Sec.~\ref{sec:fits}.

\subsection{No scaling for $G_{Ep}$ and $G_{Mp}$}

The ratio of $G_{Ep}$ and 
$G_{Mp}$ predicted by Eqs.~(\ref{eqF1}) and (\ref{eqF2})  is a simple result independent of the body form factor and dependent only on the quark form factors
\bea
R=\frac{G_{Ep}}{G_{Mp}}=3\frac{\left(f_{1+}+f_{1-}\right)-\tau \left(f_{2+}+f_{2-}\right)}
{\left(f_{1+}+5f_{1-}\right) + \left(f_{2+}+5f_{2-}\right)} \, .\label{Rratio}
\eea
This will not give scaling unless the $f_2$'s are much smaller that the $f_1$'s, but the relations (\ref{isomoments}) and (\ref{quarkmoments}) ensure that this cannot be true near $Q^2=0$.  In fact, Eq.~(\ref{Rratio}) shows that  a strong violation of scaling is  the most ``natural'' result.   In the naive case where $f_{1+}=f_{1-}=f_1$ and $f_{2+}=f_{2-}=f_2=\kappa f_1$, $R$ simplifies to
\bea
R=\frac{G_{Ep}}{G_{Mp}}=\frac{f_1-\tau f_2}
{f_1 + f_2}=\frac{1-\tau \kappa}{1+\kappa}\, , 
\label{Rsimple}
\eea
predicting that $R$ will be zero  at $Q^2\simeq2$ (GeV)$^2$ if $\kappa\simeq2$.  However, the correct asymptotic behavior for $G_E$ and $G_M$ cannot be reproduced unless the $f_2$'s go to zero faster at high $Q^2$ than the $f_1$'s, so the ratio $R$ must flatten out at high $Q^2$.   
Still, it is not hard to understand why  this simple model predicts the violation of scaling.

\subsection{Angular momentum theorem}\label{sec:j}

In this section we will use the term {\it angular momentum theorem\/}  (AMT) to  refer to the light-cone result predicting that the Pauli form factor $F_2=0$ if there are no angular momentum components in the  wave function ($L=0$).  

The derivation of the AMT is based on the view of the nucleon illustrated in Fig.~\ref{fig:one}(b). In this view the contribution shown in the figure is the matrix element of the bare quark current between the Fock component with 3 quarks and $n$ gluons (only $n=1$ is shown in the figure), which we represent by
\bea
j^\mu_n(q)=\left<P_+|3q+ng\right>\left<q|\gamma^\mu|q\right>\left<3q+ng|P_-\right>\, .
\eea
Alternatively, the contributions to the nucleon form factor can be organized as shown in Fig.~\ref{fig:one}(a), where this diagram is part of the matrix element of the {\it dressed\/} quark current (the CQ current) between the valence component of the wave function.  The equivalence of these two ideas is represented by the identity
\bea
J^\mu_I(q)=\begin {cases}{\displaystyle\sum_{n=0}^\infty j^\mu_n(q)} & {\rm light}$-${\rm front}\cr\cr
\left<P_+|3q\right>j_I^\mu\left< 3q| P_-\right>
& {\rm spectator} \end{cases} \label{lfs}
\eea
with $j_I^\mu$ the dressed quark current introduced in Eq.~(\ref{eqmiccur}).  In what follows we simplify the discussion by focusing on the anomalous moment of the quarks, related to the value of the Pauli form factors of $j_I^\mu$ at $Q^2=0$.

If the anomalous moments of the quarks are zero, Eq.~(\ref{Rratio}) predicts that $R=1$,  implying that $F_2=0$.   The nonzero value of $F_2$ in the $S$-state model comes directly from the nonzero anomalous moment of the quark, and since the light-cone formalism allows for no quark anomalous moment, there seems to be a contradiction.  

\begin{figure}
\centerline{
\mbox{
\includegraphics[width=2.0in]{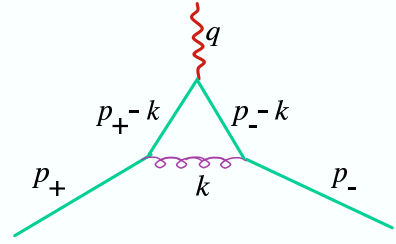} }}
\caption{\footnotesize{(Color on line) Feynman diagram for for the lowest order contribution to the CQ anomalous moment. }}
\label{fig:mu}
\end{figure}

This contradiction is easily removed by considering Fig.~\ref{fig:one} as one of the many terms contributing to Eq.~(\ref{lfs}).  In light-front language, this term is one contribution to the Fock matrix element $j_1^\mu(q)$ involving the bare quark current and the Fock state with one gluon.  In the spectator language, this term contributes to the matrix element of the lowest order anomalous magnetic moment, illustrated in Fig.~\ref{fig:mu}.  From the spectator viewpoint, this term contributes to the electromagnetic structure of the constituent quark and has nothing to do with the structure of the nucleon itself.

How exactly does the anomalous moment arise from Fig.~\ref{fig:mu})?  Since the anomalous moment is the coefficient in the $\sigma^{\mu\nu}q_\nu$ term, which is linear in $q$, it cannot be computed by evaluating the diagram at $q=0$.  Furthermore, a term linear in $q$ can emerge only from the numerator of the diagram.  Part of this numerator includes the factors
\bea
&&{\cal N}^\mu=\int d^4k \left(m_q+\not\!p_+ -\not\!k\right)\gamma^\mu  \left(m_q+\not\!p_--\not\!k\right) \qquad
\eea
where $p_{\pm}=p\pm\sfrac12 q$.  As it turns out, the term linear in $q$ comes from the momentum-dependent cross terms involving the operators $\not\!p_\pm-\not\!k$, and hence {\it require the angular-dependent\/} terms included in a quark propagator.  If this contribution is regarded as part of the wave function of the nucleon, then $L>0$ components are indeed required to generate a nonzero $F_2$, and the AMT is proved.  However, in the spectator formalism where the structure of the quark current is factored out of the wave function, $L>0$ components are not required.  

How large is the anomalous moment predicted by the diagram in Fig.~\ref{fig:mu}?  A precise calculation would require summation of all the QCD diagrams taking into account the fact that the incoming and outgoing quarks are off-shell.  To get a rough estimate, use the well know result from QED, assume that the average $\left<p_\pm^2\right>=m_q^2$, and multiply by a factor of three to include color.  This gives (remembering that we have defined $\kappa$ in nuclear magnetons)
\bea
\kappa=3\;\frac{\alpha_s}{2\pi}\frac{M}{ m_q}\, .
\eea
If we assume $\alpha_s\simeq 1$ and $M/m_q\simeq 3$, we obtain the estimate $\kappa\simeq 1.5$.  This estimate is admittedly crude, but perhaps sufficient to show that our phenomenological results for $\kappa$ are not unreasonable.

What does it matter if anomalous moment contributions are considered part of the wave function (as they are in the light-cone) or part of a CQ form factor (as they are in the spectator theory)?  As long as one recognizes where various effects are included in a calculation, it may not.  Unfortunately, our intuition is often shaped by hidden assumptions (which are usually based on our understanding of nonrelativistic quantum mechanics) so that perfectly ordinary results stated in one formalism may seem surprising when looked at from another.  Angular momentum components are one of these -- successful nonrelativistic (or semirelativistic) models based on pure $S$-state wave functions have existed for a long time, and it is important to realize that these are not ``wrong'' because they contain no angular momentum components in the wave function.   One of the objectives of this paper is to show that these nonrelativistic models can be generalized to a manifestly covariant form without losing the intuition gained from nonrelativistic physics. 

\begin{figure}
\centerline{
\mbox{
\includegraphics[width=2.0in]{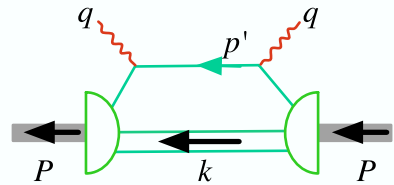} }}
\caption{\footnotesize{(Color on line) Feynman diagram for DIS.  All of the intermediate quarks are on shell. }}
\label{fig:DIS}
\end{figure}

\subsection{Where are the lower components of the wave function?}

An essential feature of this model is that the nucleon wave function, which is the product of the quark propagator and the nucleon vertex function, Eq.~(\ref{Nwf0}), is modeled by the sum of two simple, largely spin-independent scalar wave functions, as written in Eq.~(\ref{Nwf00}). 
As mentioned already in Sec.~\ref{sec:wf}, removing the quark pole from consideration is a simple way to build quark confinement into the model, but removing the spin-dependent positive energy quark projection operator from the nucleon wave function has the effect of removing the lower components of the quark spinor and replacing them by the lower components of the nucleon spinor.  Since these lower components are exactly zero when the nucleon is at rest, the bound quark must also have no lower components, and it is natural to ask what kind of interaction could produce such a result.

To answer this question, consider the covariant spectator equation that might describe the nucleon as a bound state of an off-shell quark and on on-shell ``diquark''
\bea
&&S_q^{-1}(P-p)\Psi(p,P)=
\int _k{V}(p,k;P)\Psi(k,P)\qquad \label{speq}
\eea
where $\int_k$ was defined in (\ref{intk}), $\Psi$ is the wave function defined in Eq.~(\ref{Nwf0}),  and $S_q^{-1}$ is the inverse of the quark propagator
\bea
S_q^{-1}(p_1)=m_q-\not\! p_1\, .
\eea
A kernel $V$ can be constructed that will give linear confinement in coordinate space \cite{Savkli,GM1} and the details of its construction need not be discussed here.  In the nonrelativistic limit (defined by $m_s\to\infty$) Eq.~(\ref{speq}) has the nice feature 
that it reduces to the Dirac equation
\bea
\Big(i\not\!\partial -m_q -V(r)\Big)\Psi (r)=0\, ,
\eea
where $\Psi(r)$ and $V(r)$ are three-dimensional Fourier transforms of their momentum space counterparts.

To obtain a wave function without lower components (in the rest frame) it is sufficient to replace the  kernel by
\bea
{\cal V}(p,k;P)= {\cal O}(p,P)V(p,k;P)\overline{{\cal O}} (k,P) \label{vtov}
\eea
where ${\cal O}(p,P)=S^{-1}_q(p_1)[M+\not\!\!P]$, $p_1=P-p$, and for any Dirac operator $\overline{A}=\gamma^0A^\dagger\gamma^0$.  The construction preserves the hermiticity property of the kernel, $V=\overline{V}$.   With this substitution the wave function satisfies a new spectator equation
\bea
&&\Psi(p,P)=[M+\not\!P]
\int _k{\cal V}(p,k;P)\overline{{\cal O}}(k,P)\Psi(k,P)\qquad \label{speq2} \qquad
\eea
showing immediately that $\Psi$ satisfies the Dirac equation (\ref{Dirac}), sufficient to ensure that its lower components are zero in the nucleon rest frame.  

This demonstration shows precisely how the features of the spectator theory can be exploited to build in this phenomenology: since the quark (in this case) is off-mass-shell, and since the kernel $V(p,k;P)$ always depends on the total four-momentum, the substitution (\ref{vtov}) does not violate any of the constraints that must be satisfied in constructing a phenomenological kernel.

\section{Deep inelastic scattering}

Before studying the details of the predictions for the form factors, we will show how this model is able to describe deep inelastic scattering (DIS).

\subsection{Quark distribution function} \label{sec:qdis}

The DIS cross section can be calculated from the imaginary part of the forward handbag diagram shown in Fig.~\ref{fig:DIS}.  The cross section depends on the hadronic tensor
\begin{widetext}
\bea
W_{\mu\nu}(q,P)&=&\frac{3}{2S+1}\sum_{s,s_1,s_2}\int\int \frac{d^3p'\,d^3 k}{(2\pi)^6 4e_qE_s}
(2\pi)^4\delta^4(p'+k-q-P)J^\dagger_\nu J_\mu\cr
&\equiv&-4\pi M\left\{\left(g_{\mu\nu}-\frac{q_\mu q_\nu}{q^2}\right)W_1-\left( P_\mu -
\frac{P\cdot q\,q_\mu}{q^2}\right)
\left( P_\nu -
\frac{P\cdot q\,q_\nu}{q^2}\right) \frac{W_2}{M^2}\right\}\label{hadcurr}
\eea
\end{widetext}
where $e_q=\sqrt{m_q^2+{\bf p}'^2}$ is the energy of the on-shell quark in the final state, $E_s$ is the diquark energy encountered before,  $W_1$ and $W_2$ are the DIS structure functions, and $J_\mu$ is the hadronic current
\bea
J_\mu=-\bar u({\bf p}',s_1) j_\mu(p',p)\Psi_N(P,k) \label{DIScurrent}
\eea
with $j_\mu(p',p)$ the quark current defined in Eq.~(\ref{eqmiccur}) and $\Psi_N$ the nucleon wave function defined in Eq.~(\ref{psiRel}).  Here the isospin $I$ and the polarizations of the diquark and nucleon, $s_2$ and $s$, have been suppressed.

Detailed evaluation of the hadronic tensor is sketched in Appendix A.  After the spin sums have been carried out, the structure functions become
\bea
W_1&=& \frac{\lambda^2\,e^2_I}{2\pi M} \int\!\!\!\! \int_{p'k} 
(P\cdot p')\;\psi^2(P,k)\cr
W_2&=&\frac{\lambda^2\,e^2_I}{2\pi M}  \int\!\!\!\! \int_{p'k}
2M^2A\;\psi^2(P,k) \, , \label{eq:W12}
\eea
where $\lambda$ was defined in Eq.~(\ref{asy}), the isospin-dependent charge operator is
\be
e_I^2=\sfrac56 + \sfrac16\,\tau_3=\begin{cases} 2e_u^2+e_d^2 & $proton$\cr
e_u^2+2e_d^2 & $neutron$\end{cases}\, , \label{eq:ei}
\ee
the integral is
\bea
\int\!\!\!\! \int_{p'k}
&\equiv& \int\!\!\!\int\frac{d^3p' d^3k}{(2\pi)^2\,4e_qE_s}\delta^4(p'+k-P-q)\cr
&=&\int\frac{d^4k}{(2\pi)^2}\delta_+(m_q^2-p'^2) \delta_+(m_s^2-k^2)\, , \label{DISint}
\eea
and
\bea
A=\frac{(p'\cdot P) Q^2+(P\cdot q)(p'\cdot q)}{M^2Q^2+(P\cdot q)^2}\, . \label{eq:A}
\eea

The expressions for $W_1$ and $W_2$ are covariant, but it is convenient to evaluate them in the laboratory system using light-cone coordinates. In our notation, an arbitrary four-vector $v$ is written
\bea
v=\{v_+,v_-,{\bf v}_\perp\}\qquad
v_\pm = v_0\pm v_3
\eea
so that the scalar product is
\bea
v\cdot u = v_\mu u^\mu=\sfrac12(v_+u_-+v_-u_+)-{\bf v}_\perp\cdot{\bf u}_\perp
\eea
The four vectors in the laboratory frame are
\bea
P&=&\left\{M,0,0,0\right\}\cr
q&=&\left\{\frac{Q^2}{2Mx},0,0,\sqrt{Q^2+\frac{Q^4}{4M^2x^2}}\right\}
\eea
where $x$ is the usual Bjorken scaling variable.  In the DIS limit, in light-cone form, these momenta become
\bea
P&=&\{M,M,{\bf 0}\}\cr
q&=&\left\{\frac{Q^2}{Mx}, -Mx, {\bf 0}\right\}\, .
\eea

Using this notation, we find that the structure functions $W_1$ and $W_2$ scale, and satisfy the Callen-Gross relation
\bea
\nu W_2 = 2MxW_1\equiv e_I^2\,x f_q(x)\, .
\eea
with a quark distribution function $f_q(x)$ given by 
 \bea
 f_q(x)=\frac{\lambda^2}{4\pi } \int\frac{d^2k_\perp}{(2\pi)^2(1-x)} \psi^2(\chi_{\rm DIS}) \label{kpint}
 \eea
where $\chi_{\rm DIS}$ is the value of $\chi$, Eq.~(\ref{chi}), in the DIS limit.  If
\be
r=\frac{m_s}{M} \label{ratio}
\ee 
and $k_\perp^2=Mm_sy$, then
\bea
\chi_{\rm DIS}= -2+\frac {r+y}{1-x} +\frac{(1-x)}{r}\, .
\eea
The values of $k_\pm$ were fixed by the $\delta$ function restraints in Eq.~(\ref{DISint}), leaving an integral only over the perpendicular components of $k$.  Using the wave function (\ref{eqpsi0}) the $k_\perp$ integral is easily evaluated, giving
\bea
f_q(x)&=&\frac{\lambda^2\,N_0^2}{16\pi^2 r(\beta_1-\beta_2)^2}H\, ,\label{quarkfx}
\qquad
\eea
where
\bea
H&=&\frac{1}{R_1}+\frac{1}{R_2}-\frac{2}{\beta_2-\beta_1}\log \frac{R_2}{R_1}\nonumber\\
R_i&=&\beta_i-2+\frac{r}{1-x}+\frac{1-x}{r}\nonumber\\
&=&\frac{x}{Mm_s}\left[\frac{m_{\beta i}^2}{x}+\frac{m_s^2}{1-x}-M^2\right]
\eea
and $m_{\beta i}^2=Mm_s\beta_i+(M-m_s)^2$.

Before we examine the implication of Eq.~(\ref{quarkfx}), we study the normalization of $f_q(x)$.

\subsection{Normalization of the quark distribution function} \label{sec:norm}
 
 The normalization condition (\ref{norm}) for the nucleon wave function can be expressed as a normalization condition for the quark distribution function.  Writing the normalization integral in terms of light-cone variables
 \bea
1&=& \int\frac{dk_+\, dk_-\, d^2k_\perp}{2(2\pi)^3}\delta_+(m_s^2-k^2)\psi^2(P,k)\nonumber\\
&=&\int_{0}^\infty \frac{dk_-}{k_-}\int \frac{d^2k_\perp} {2(2\pi)^3}\psi^2\left(k_+=\frac{m_s^2+k_\perp^2}{k_-}\right)\, .\qquad\label{norm1}
 \eea
Multiplying both sides of (\ref{norm1}) by $\lambda^2$, defining $k_-=M(1-x)$, and using (\ref{quarkfx}) gives the relation
\bea
\lambda^2=\int_{-\infty}^1dx\, f_q(x)\qquad \mbox {(spectator theory)}\, . \label{norm2}
\eea

The contributions to this normalization integral from the region $x<0$ are a feature of the spectator theory and arise because, by design, the spectator theory includes only contributions from the singularities of the spectator \cite{Gross:1983yt}.     If additional singularities in the wave function are also taken into account, the contribution to the norm (\ref{norm1}) from the region of $x$ between $-\infty$ and 0 will be canceled.  In Appendix B we show that including the singularities of the wave function limits the $k_-$ integral to the region $[0,M]$, giving
\bea
\lambda^2=\int_{0}^1dx\, f_q(x)\qquad \mbox{(light-front)}\, . \label{norm3}
\eea
Hence both theories (spectator and light-front) give the same quark distribution amplitude; the only difference is the normalization condition (\ref{norm2}) or (\ref{norm3}).

To compare with experiment, we use the valence distributions obtained from a recent global fit \cite{Martin:2002dr}.  Results will be compared to the proton average 
\bea
xf_q(x)|_{\rm expt}=\sfrac49 xu_V+\sfrac19 xd_V
\eea
(where $xu_V$ and $xd_V$ are Eqs.~(1) and (2) of Ref.~\cite{Martin:2002dr}) and is normalized to
\bea
\int_0^1 dx f_q(x)|_{\rm expt}&=&1.006\simeq1.00\nonumber\\
\int_0^1 dx x f_q(x)|_{\rm expt}&=&0.171\, .
\eea
For the light-front, consistency therefore requires choosing $\lambda=1$, while for the spectator we must have $\lambda>1$ so the integral over the physical range $0\le x \le1$ will be exactly unity.

We now turn to a discussion of the fits to the form factors and the quark distribution functions.

\begin{figure*}
\centerline{
\mbox{
\includegraphics[width=6in]{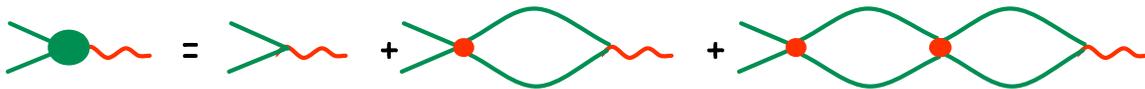} 
}}
\caption{\footnotesize{(Color on line) The first few terms of the bubble sum for the quark form factors.  The first term is the bare charge, missing from the $f_2$ form factors.}}
\label{fig:bubble}
\end{figure*}

\section{Results}\label{sec:fits}

\subsection{Vector dominance model for the quark form factors}

Motivated by the vector dominance model (VDM), the quark form factors are written in the following form
\bea
f_{1\pm}(Q^2)&=&\lambda + \frac{(1-\lambda)}{1+Q_0^2/m_v^2} +\frac{c_\pm\,Q_0^2/M_h^2}{(1+Q_0^2/M_h^2)^2}\nonumber\\
f_{2\pm}(Q^2)&=&\kappa_{\pm}\left\{
\frac{d_\pm}{1+Q_0^2/m_v^2} +\frac{(1-d_\pm)}{1+Q_0^2/M_h^2} \right\}\, ,\qquad
\label{VDM}
\eea
where $m_v$ is the $\rho$ ($\omega$) mass for the isovector (isoscalar) form factors, and $M_h=2$ is a fixed heavy mass intended to approximate the contributions from all heavy vector mesons in each channel (this could depend on the isospin but this possible dependence is ignored in these fits).  All masses and momenta are in units of the nucleon mass $M$, so that, for example, $Q_0^2=Q^2/M^2$.  

It may be helpful to give a brief justification for the quark form factor models (\ref{VDM}).  From a microscopic point of view, the VDM for the form factors emerges from a sum similar to that shown in Fig.~\ref{fig:bubble}.  Here,  the quark-quark interaction is approximated by a contact term,   giving
\bea
f_{1}&=&\lambda e + gB(s)\lambda e + [gB(s)]^2\lambda e +\cdots\nonumber\\
&=& \lambda e + \frac{g\lambda e B(s)}{1-gB(s)}
\eea
Choosing a suitable $B(s)$ that goes like $Q^{-2}$ as $Q^2\to\infty$ then gives Eq.~(\ref{VDM}) for the form factors $f_{1\pm}$.  The construction of $f_2$ is similar, with the first term missing because there is no bare anomalous moment at infinite $Q^2$.

\subsection{Fitting procedure}

 During each step in the fit, we first normalized the wave function using Eq.~(\ref{norm}), determined the anomalous moments of the quarks by fixing the proton and neutron magnetic moments, and then determined the asymptotic charge parameter $\lambda$ and diquark mass $m_s$ by satisfying the following two constraints
\bea
\left<f\right>&\equiv&\int_0^1 dx f_q(x) =1\nonumber\\
\left<xf\right>&\equiv&\int_0^1 dx x f_q(x) =0.171
\label{consistency}
\eea
where the quark distribution amplitude $f_q(x)$ was defined in Eq.~(\ref{kpint}).  Even though the form factors are independent of $m_s$ (recall that it was scaled out of the results) the deep inelastic quantities are not. Since $f_q$ includes $\lambda$ in its definition, $m_s$ was first determined from the ratio $\left<xf\right>/ \left<f\right>$, and then $\lambda$ was fixed by the normalization requirement $\left<f\right>=1$.  This procedure was quite stable and led quickly to a good determination of the other parameters.

In this paper we  chose to define the charge of the $u$ quark (for example) to be 2/3 at $Q^2=0$, and to normalize the wave function to unity.  QCD suggests that it might be preferable  to define the $u$ quark charge to be 2/3 at $Q^2=\infty$.  This can be done by dividing the quark form factors by $\lambda$ which changes the charge at $Q^2=0$  to $2/(3\lambda)$.  If the wave function is then normalized to $\lambda$ (instead of unity, as is now the case), all of the formula for the form factors will be unchanged.  However, the numerical results will change slightly because the normalization requirement (\ref{consistency}) will change to $\left<f\right>=\lambda$ (keeping the definition of  $f_q$ unchanged).  This will give the same value of $m_s$, but a slightly different value of $\lambda$, leading to a different minimum with slightly different numerical results.  We save discussion of this subtlety for future work. 

\begin{table}
\begin{center}
\begin{tabular}{rc l l l c c r }
Model &$\beta_1, \beta_2$ & &
$c_+, c_-$ &
$d_+, d_-$ &
$b_E, b_M$ &
$\lambda, r$ &
$N_0^2, \chi^2$  \\
\hline
I(4)&0.057 &     & 2.06 & $-0.444$  
& $--$ &  1.22    & 10.87\\
       &0.654   &  &2.06$^*$  & $-0.444^*$  
 & $--$ & 0.88  & 9.26  \\
\hline
II(5)&  0.049   & & 4.16  & $-0.686$     
& $--$  &   1.21 &    11.27\\
        &  0.717   &  & 1.16 &   $-0.686^*$   
 & $--$ & 0.87 &    1.36 \\
\hline
III(6)&  0.078   & $\;$ & 1.91   & $-0.319$  
& 0.163   &   1.27  &  12.36  \\
        &  0.598    &  & 1.91$^*$  &  $-0.319^*$
&  0.311 & 0.89     & 1.85 \\
\hline
IV(9) &  0.086   &  & 4.48    & $-0.134$  &   0.079 &     1.25 &  8.46  \\
&  0.443    &  & 2.45  & $-0.513$  & 0.259  &  0.89   & 1.03 \\
\hline
\end{tabular}
\end{center}
\caption{The two lines for every model give the values of the 8 possible (9 for Model IV) 
adjustable parameters and the 4 constants fixed by the constraints.  The fixed constants are $\lambda$, the asymptotic value of $f_1$ given in Eq.~(\ref{asy}), the diquark mass ratio $r$, Eq.~(\ref{ratio}),  the normalization parameter $N_0^2$, and the $\chi^2$ per data point.  In each case $\kappa_+ = 1.639$ and $\kappa_-=1.823$.  Parameters labeled with an $^*$ were constrained during the fit to equal the one above it.  For Model IV the heavy mass defined in Eq.~(\ref{VDM}) was also adjusted, giving a best value of $M_h=2.556$.}
\label{table}
\end{table}

\begin{table}
\begin{minipage}{3.5in}
\begin{tabular}{l rrrrr }
&$\quad$ I $\quad$&$\quad$ II $\quad$&$\quad$III $\quad$& $\quad$IV$\quad$& $\quad$expt $\quad$\\
\hline
$r_p^2$ &0.764&0.791&0.851&0.703& 0.780(25)\\
$r_n^2$ &$-0.005$&$-0.104$&$-0.241$&$-0.102$& $-0.113(7)\;$ \\
$\chi^2_{\rm radii}$ &118$\;$&0.89$\;$ &170$\;$  &6.11$\;$ & \\ [2pt]
\hline
\end{tabular}
\caption{The values of the charge radii and the $\chi^2_{\rm radii}$ for each model.}
\label{table2}
\end{minipage}
\end{table}

\begin{figure*}
\centerline{
\mbox{
\includegraphics[width=6in]{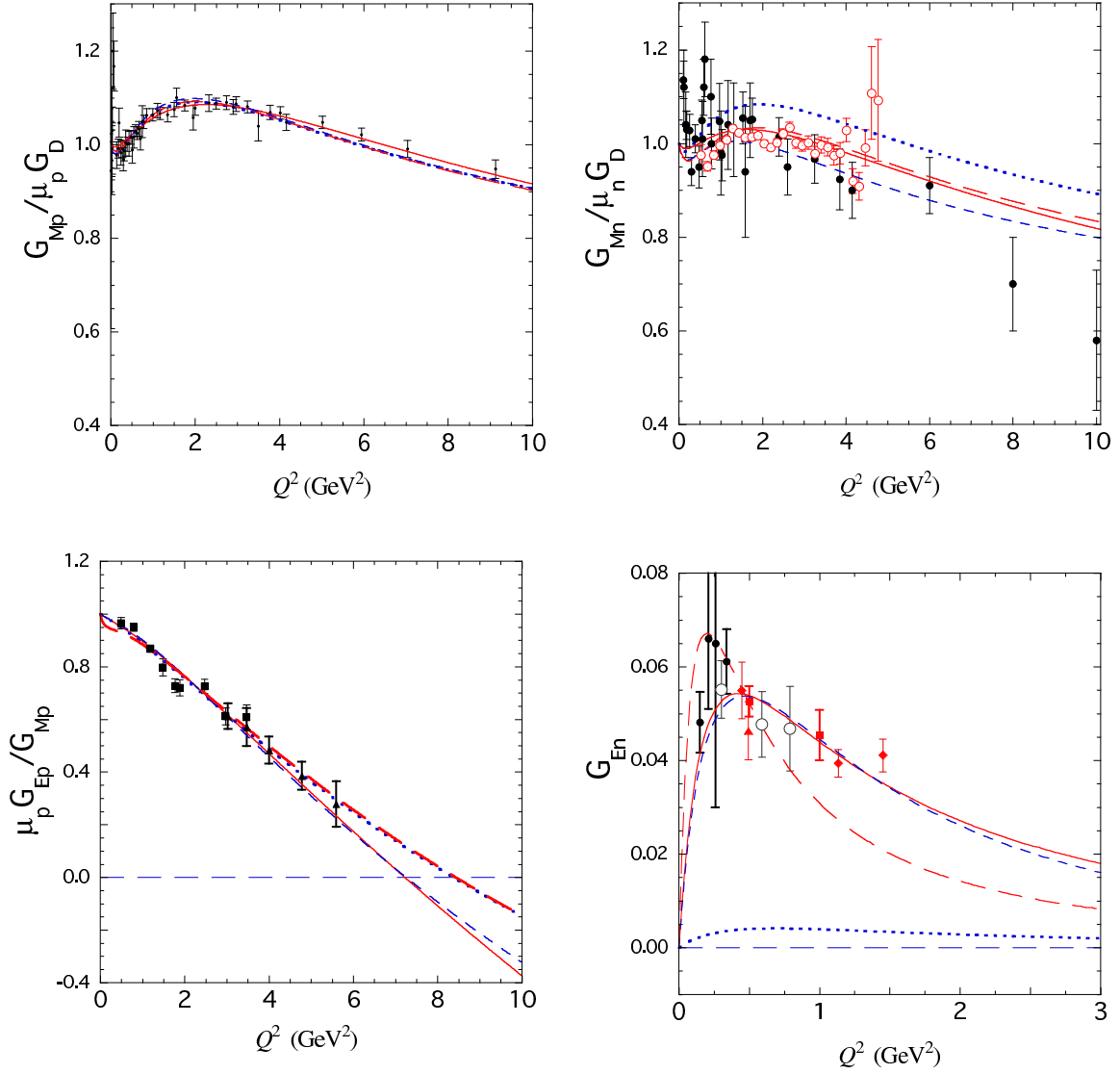} 
}}
\caption{\footnotesize{(Color on line) Data for the nucleon form factors compared with the four models discussed in the text: Model I (dotted line), II (short dashed line), III (long dashed line) and IV (solid line). The $G_{Mp}$ data is from Arrington  \cite{Arringtonpri}, and the $G_{Mn}$ data includes that used by Bosted (solid circles) in his global fits to the form factors \cite{Bosted94} and the unpublished data from JLab Hall B (open circles) \cite{Brooks:2005gt}.  The $G_{Ep}$ data are from JLab Hall A, Jones \etal  and Punjabi \etal \cite{Jones99,Punjabi05} (squares) and  Gayou, \etal \cite{Gayou01} (triangles).  The $G_{En}$ data are single $Q^2$ points from MAMI \cite{Herberg99,Ostrick99}, NIKHEF \cite{Passchier99}, and MIT-Bates \cite{Eden94} (solid circles),  and from JLab Hall C by Zhu, \etal \cite{Zhu} (triangle), Warren, \etal  \cite{Warren} (squares), and Madey, \etal  \cite{Madey03}(diamonds), and from MAMI by Glazier \cite{Glazier04} (open  circles).  Only $G_{En}$ data obtained from deuteron targets are included.
For a list of the data see the
nucleon form factor data base \cite{database}.}}
\label{fig1}
\end{figure*}

\begin{figure*}
\centerline{
\mbox{
\includegraphics[width=6in]{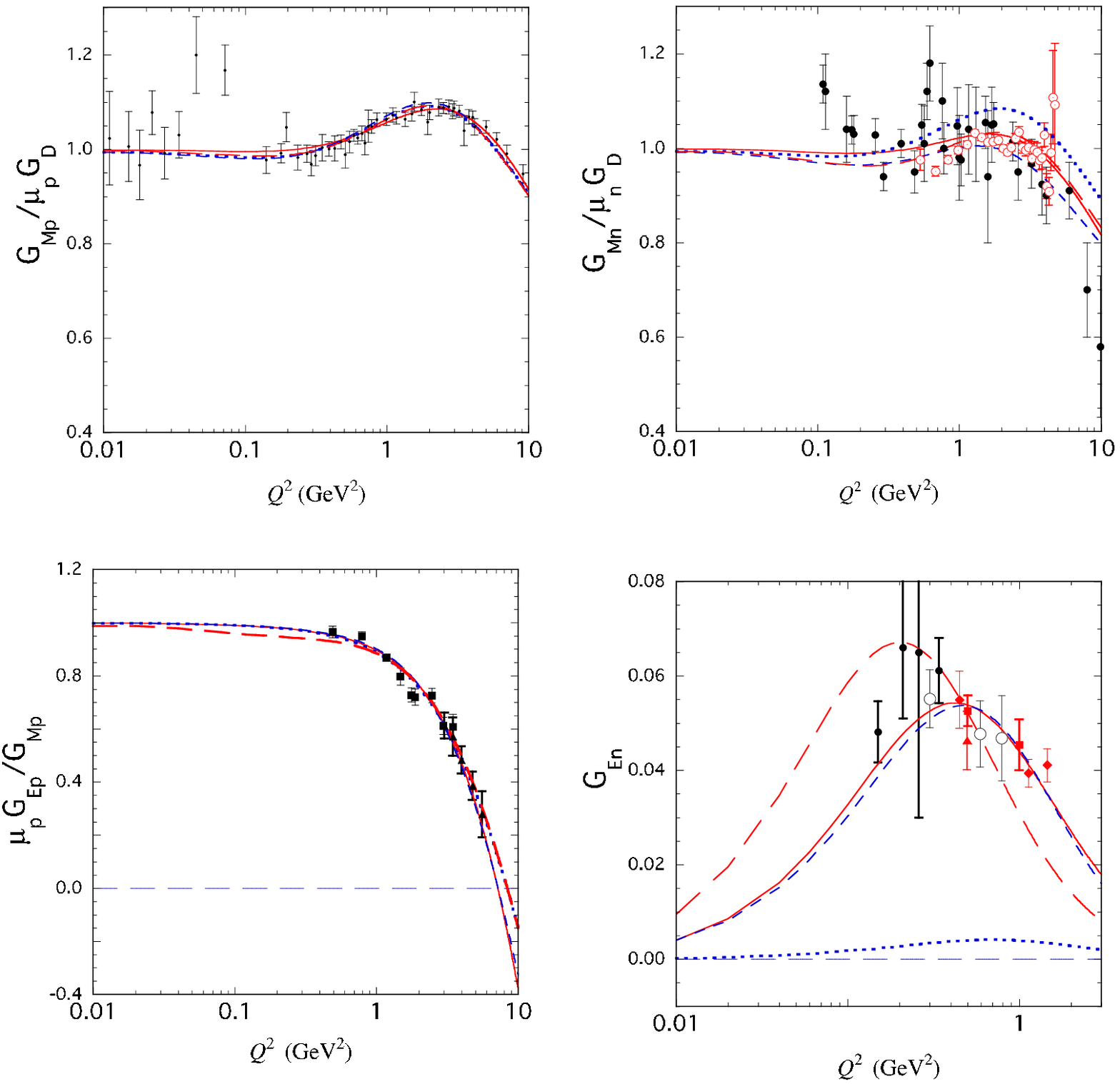} 
}}
\caption{\footnotesize{(Color on line) Log plot of the same curves and data shown in Fig.~\ref{fig1}.  [For an explanation see the caption to Fig.~\ref{fig1}.]  These curves show the low $Q^2$ structure discussed in Ref.~\cite{Friedrich:2003iz}.}}
\label{fig1a}
\end{figure*}

\subsection{Fits without a two-pion cut}

Since the VDM for the form factors includes a $\rho$ pole and the $\rho$ couples to the $2\pi$ channel, the quark form factor includes two-pion terms.  It is therefore next to impossible to separate out a pion cloud term from the quark form factor contributions.    However, the two-pion channel contributes a cut starting at $q^2=-Q^2>4 m_\pi^2$, and it has been known for many years that the singularity associated with the onset of this cut, which goes like $(q^2-4m_\pi^2)^{3/2}$, has a strong effect on the form factors near $Q^2=0$.  The same cannot be said for the isoscalar form factors. The isoscalar singularity is produced by the onset of a three-pion continuum, with a three-body phase space starting smoothly at $9m_\pi^2$. It is much weaker, more distant, and less important. 

The VDM model described in Eq.~(\ref{VDM}) does not have the two-pion singularity.  In the next section we will discuss how this singularity can be added, but here we look at the consequences of ignoring this singularity.

We present four models with different features that represent different aspects of the physics.   The parameters for each model are shown in Table \ref{table}.  In this section we discuss the first two of these models, I and II.  These models do not contain the pion-cut singularity.  Model I is simplest model with the fewest parameters (4).  It assumes approximate isospin symmetry by constraining $f_{1+}\simeq f_{1-}$ and $f_{2+}\simeq f_{2-}$ (they are not equal because  $m_\rho \ne m_\omega$ and $\kappa_+\ne \kappa_-$).  As already discussed in Sec.~\ref{sec:nff}, approximate isospin symmetry for a pure $S$-wave model of the nucleon gives a very small neutron electric form factor, and this is show in Fig.~\ref{fig1} and Table \ref{table2} (showing that the charge radius of the neutron is off by almost two orders of magnitude).  We do not believe that Model I gives a credible description of the form factors, even though its predictions are very close to many other simple quark model calculations.  We present it to show what can be achieved from very few assumptions.

As discussed above, one of the ways to obtain a reasonable neutron charge form factor is to break the isospin symmetry.  Model II  allows the isospin to be maximally broken by the $f_1$ form factors, but preserves the approximate isospin symmetry of the $f_2$ form factors.  This adds only one more parameter, but does an excellent job of fitting the data, and incidentally, gives almost perfect results for the two charge radii.  This model is perhaps the best compromise between the need to obtain a good description of the form factors, and the  phenomenological requirement that the description be economical, using only a few parameters.
  
Now we discuss the effect of adding a term that explicitly includes the two-pion threshold singularity. 

\subsection{Including a pion cloud (two-pion cut)}

\begin{figure}
\centerline{
\mbox{
\includegraphics[width=3in]{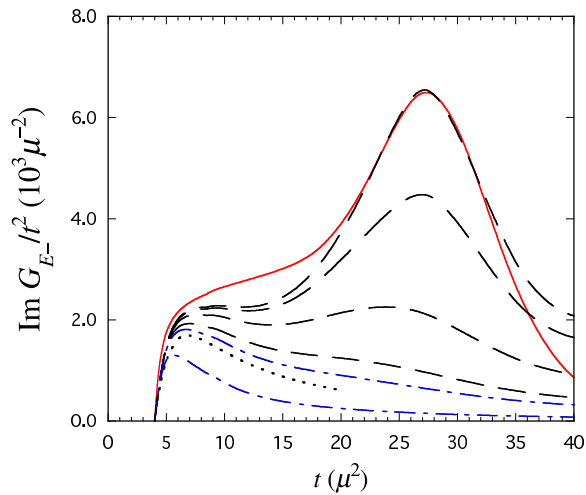} }}
\caption{\footnotesize{(Color on line) Im $G_{E-}/t^2$ compared to the experimental fit of Ref.~\cite{Hohler:1974eq} (upper solid curve) and the one loop ChPT calculation of Ref.~\cite{Kaiser:2003qp} (lower dotted curve ending at $t=20$).  The units are pion masses, $\mu=m_\pi$.  The dashed lines, in order of decreasing size, are the models with values of $b_E =$ 0.55, 0.45, 0.30, and 0.20.   The two dot dashed lines correspond to Model III (upper) and Model IV (lower).}}
\label{fig:imE}
\end{figure}

\begin{figure}
\centerline{
\mbox{
\includegraphics[width=3in]{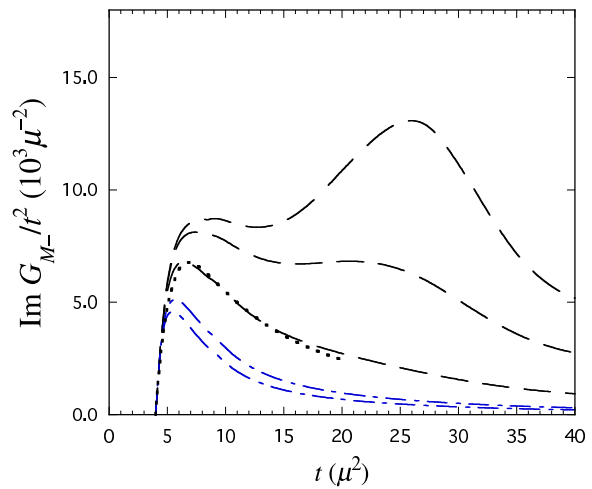} }}
\caption{\footnotesize{(Color on line) Im $G_{M-}/t^2$ compared to one loop ChPT calculation of Ref.~\cite{Kaiser:2003qp} (lower dotted curve ending at $t=20$).  The dashed lines, in order of decreasing size, are the models with values of $b_M =$ 1.50, 1.00, and 0.55.  The two dot dashed lines correspond to Model III (upper) and Model IV (lower). }}
\label{fig:imM}
\end{figure}

The imaginary part of the isovector form factors has been calculated in chiral perturbation theory (ChPT) \cite{Kaiser:2003qp,Hammer:2003qv}, and also extracted from data \cite{Hohler:1974eq}.  We model this behavior by adding extra terms to $G_{E-}$ and $G_{M-}$ of the form
\bea
\Delta G_{i-}(Q^2)=\frac{b_i^2}{b_i\left(1+\frac{Q^2}{m_1^2}\right)\left(1+\frac{Q^2}{m_2^2}\right)+a_i\frac{\Gamma(Q^2)}{Q^2}} \label{dGcloud}
\eea
where the width function is
\be
\Gamma(Q^2)=\frac{\mu^2(4+Q^2/\mu^2)^{3/2}}{\left(1+\frac{Q^2}{2\mu^2}\right)}
\ee
The choice of this function is discussed in some detail in Appendix C.  The parameters $a_i$, $m_1$ and $m_2$ were chosen to reproduce the threshold behavior of the imaginary parts of the isovector $G_E$ and $G_M$ form factors, calculated in chiral perturbation theory, and, for $G_E$, also obtained from experiment.  The imaginary part near threshold is insensitive to the value of the additional parameter $b_i$ adjusted during the fit.  The behavior of the imaginary parts is shown in Figs.\ \ref{fig:imE} and \ref{fig:imM}.  

Models III and IV both include the pion cloud term (\ref{dGcloud}).  Model III shows that adding this term significantly improves the description even when the approximate isospin symmetry used in Model I is still maintained.  We conclude that we may significantly improve the fit to the data by {\it either\/} breaking the symmetry of $f_1$ {\it or\/} by adding a pion cloud term.   However, these two alternatives are not completely equivalent.  Breaking the symmetry of $f_1$ (Model II) produces a better fit (lower $\chi^2$), much better radii, and also gives a $G_{En}$ with a  better shape.  The shape of $G_{En}$ seems to be particularly sensitive to how the isospin symmetry is broken.    

Finally, Model IV shows that a combination of the two effects 
studied separately in Models II and III can give a precision fit to the data. 
[To obtain this the heavy mass $M_h$ was also varied, 
adding one more parameter.]  
The expected $\chi^2$ for a statistically perfect fit should lie in the range $\chi^2/n=1\pm\sqrt{2/n}$, where $n$ is the number of data.  Here the number of data is 117, leading to an expected $\chi^2/n=1\pm 0.13$ \cite{Ber88}, nicely consistent with our result of 1.03, or 1.12 if the radii are included in the fit. One of the reasons a high quality fit is possible is that we used data for $G_{Mp}$ recently reanalyzed by Arrington  \cite{Arringtonpri}; these data are more consistent with the new $G_{Ep}$ data, and this consistency is required for a good fit.  The new JLab high precision $G_{Mn}$ data are not yet final \cite{Brooks:2005gt} and were therefore  not included in our fit; if the final data set  remains close to the preliminary one a refit of the model may be necessary, and we have not investigated the effect it will have. The model predicts that $G_{Ep}$  vanishes
for $Q^2\sim$7.5 GeV$^2$.

\subsection{Quark distribution functions}

\begin{figure}
\centerline{
\mbox{
\includegraphics[width=3in]{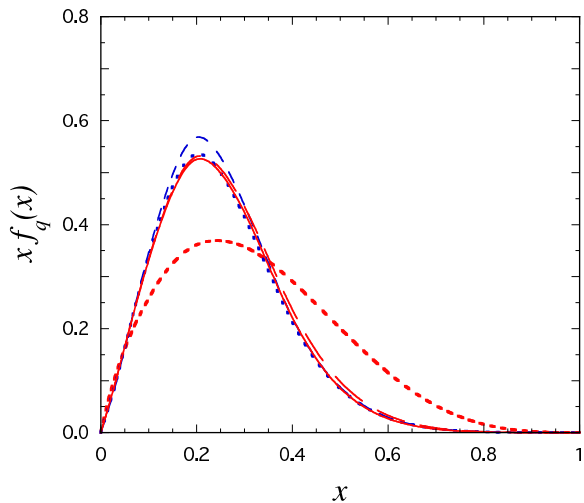} 
}}
\caption{\footnotesize{
The valence quark distribution $xf_q(x)$ as a function of $x$.  
Each curve is normalized to 0.171, the experimental momentum fraction carried by valence quarks in the proton.  The thick short dashed line is the empirical fit from 
Ref.~\cite{Martin:2002dr} (evolved to $Q^2=1$ GeV$^2$), and the other lines (with the same line style used in Fig.~\ref{fig1} and nearly indistinguishable from one another) are the four models discussed in the text. }}
\label{fig:quark}
\end{figure}

The results for the quark distribution function obtained from the four models are shown in Fig.~\ref{fig:quark}.  Note that all the models give very similar results; the distributions are all too narrow and approach zero as $x\to 1$ too fast, but at least give a qualitatively reasonable  description.  Better agreement would have been obtained if we had compared the models with a quark distribution function evolved to a higher $Q^2$ (4 GeV$^2$, for example), which would be somewhat more peaked than the one shown in the figure (evolved to $Q^2=1$ GeV$^2$).  We have made no attempt here to improve the agreement between our result and experimentally determined distributions; this is a subject for future work.       

\begin{figure}
\centerline{
\mbox{
\includegraphics[width=3in]{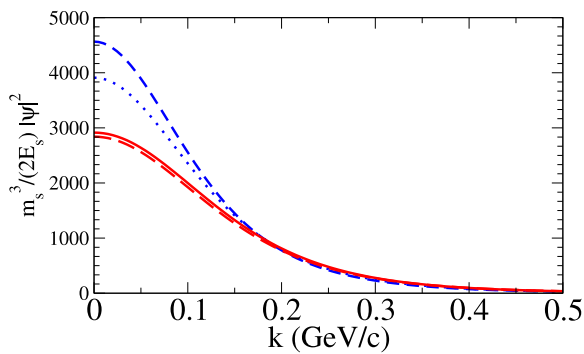} 
}}
\caption{\footnotesize{
The distribution $\rho(k)=m_s^3|\psi_0(\tilde P,k)|^2/(2E_s)$ as a function of the momentum  $k$ for each of the four models discussed in the text (drawn with the same line style used in Fig.~\ref{fig1}).  This distribution is normalized to $4\pi\int_0^\infty k^2 dk\, \rho(k)/(2\pi m_s)^3=1$.}}
\label{fig_rho}
\end{figure}

Finally, Fig.~\ref{fig_rho} shows the normalized momentum distribution for the nucleon in its rest frame (where it is spherical).  The larger size of the wave functions for Models I and II are compensated by a slightly smaller size at larger $k$; the difference is exaggerated in the figure because the  factor $k^2$, included in the normalization integral, is excluded from the distribution.

\section{Summary and Conclusions}

\subsection{Overview and comparison with previous work}

This work presents a simple covariant model for the nucleon based on the following assumptions:   (i) the nucleon is composed of 
three valence constituent quarks (CQ, massive extended particles dressed by the quark-antiquark interaction, the pion cloud, and  gluon sea, all parametrized by quark form factors), 
(ii) the three-quark system is described by an {\it internal\/} wave function consistent with the properly symmetrized covariant spectator formalism \cite{Gross69}, and the overall center-of-mass motion of the total system is described by a free Dirac equation for a particle of mass $M$ (the nucleon mass), (iii) the internal wave function has a structure built entirely from $S$-wave components with exactly the same spin-isospin content as the simplest nonrelativistic $SU(2)\times SU(2)$ model, (iv) the CQ form factors are normalized to the quark charges in the confinement limit ($Q^2=0$) and reproduce the behavior of pointlike quarks in the large $Q^2$ regime giving form factors that behave (up to logarithms) like pQCD  at very large $Q^2$, and (v) the wave function has the appropriate nonrelativistic limit.   A nice feature of this model is that the two components of the wave function, corresponding to spin-0 and spin-1
diquark states, are described
by one scalar wave function, as required by symmetry under quark interchange.

Aspects of this model have been included in previous work.  The use of CQ form factors is certainly not new, dating back to the early valon model \cite{Hwa:1980tj}, and has been pursued in many more recent papers \cite{Cardarelli:1995ap,DeSanctis05}.  
Others have used a vector dominance model to describe the form factors, with an additional phenomenological factor added to simulate the body form factor \cite{Bijker04,Bijker05,Gustafsson06}.  The two-pion cut is also included in this work \cite{Bijker04}.  
Following different lines, pQCD inspired calculations have
appeared \cite{Belitsky02,Kaskulov03}, 
as well as other calculations 
based on generalized parton distributions \cite{Stoler01}, 
QCD sum rules \cite{Braun06} and  
lattice QCD \cite{Matevosyan05}.  
A comprehensive  
review can be found in Refs.\  
\cite{Punjabi05,HWright04,Arrington06,Perdrisat06}.

The fits presented in this paper were obtained by first adjusting the quark anomalous moments so that the magnetic moments of both the neutron and proton are reproduced {\it exactly\/},  and then adjusting the ``diquark'' mass and the asymptotic quark charge to give the correct experimental result  for the number and total momentum fraction carried by {\it valence\/} quarks in DIS.  With aspects of both the low and high energy behavior fixed, it is then possible to study the sensitivity of the form factors to various physical assumptions.

The physics is illustrated using four models of increasing complexity.  Model I examines the consequences of the simplest assumption: the quark form factors are approximately independent of isospin (some breaking is built in because  $m_\rho\ne m_\omega$ and $\kappa_+\ne \kappa_-$).    With only four parameters (two in the wave function and one each for $f_1$ and $f_2$) the dotted lines shown in Figs.~\ref{fig1} and \ref{fig1a} are obtained, giving a good description of all the form factors except for the neutron charge form factor, $G_{En}$.  Similar results were obtained in recent work by Juli\'a-D\'iaz, Riska, and Coester \cite{Julia-Diaz:2003gq}.  Note that the nonscaling behavior of the proton charge form factor is easily described.  Ironically, $G_{En}$ is far more important to our understanding than the more popular $G_{Ep}$ measurements.   

To describe the neutron charge form factor in a pure $S$-wave model it is necessary to break the isospin invariance.  There are two ways to do this: split at least one of the quark form factors ($f_{1+}\ne f_{1-}$ or $f_{2+}\ne f_{2-}$), or add a term describing the two-pion cut (starting at $t=q^2=-Q^2>4m_\pi^2$).  These two methods are, to some extent, merely different ways to model the same physics: the two-pion cut coming from the  $\rho$ contribution (isospin=1) is much stronger  and has more structure at low $t$ than the corresponding three-pion cut associated with the $\omega$ contribution (isospin=0).  This mechanism automatically breaks isospin symmetry.  This feature has been included in the recent vector dominance models of Bijker \etal \cite{Bijker04}.

We found that breaking the $f_1$ symmetry was far more efficient than breaking the $f_2$ symmetry, and Model II shows the dramatic effect of allowing $c_+$ and $c_-$ [defined in Eq.~(\ref{VDM})] to vary independently.

Model III achieves a similar result, but its charge radii are quite poor, and the $G_{En}$ given by this model has the wrong shape.

Finally, Model IV shows that we can obtain a precision fit if both the symmetry of $f_1$ and $f_2$ is broken, and the pion-cut term is added.

Comparing Models II, III, and IV, 
we conclude that we can fit the data by  
including an explicit pion cloud and/or adjusting the  vector dominance contribution (coefficients $c_\pm$).  This result is consistent with Ref.~\cite{Meissner07},  which states that
``an unambiguous extraction of the pion cloud 
contributions is not possible.''
Furthermore, the pion cloud descriptions of  Refs.~\cite{Kaiser:2003qp,Hammer:2003qv} do not determine the real parts of the form factors, except possibly in  the limited region 
$Q^2< 0.1$ GeV$^2$ (as noted in Ref.~\cite{Wang07}).  In any case, our model for the pion cloud contribution for the nucleon form factor 
in the spacelike region gives a contribution that differs only slightly from the charge (or magnetic moment) in the small  $Q^2$ region, and decreases 
quickly with increasing $Q^2$.  We conclude that we can describe the nucleon form factor data 
using vector dominance only.

In all the models we have very different values for 
the parameters $\beta_1$ and $\beta_2$ (one order of magnitude).
This is evidence of an almost perfect separation,
in dynamics, between the low momentum 
($\beta_1$) and the high momentum 
 ($\beta_2$) regimes.

The overall success of this family of models, each of which is consistent with  the expected  pQCD asymptotic behavior (up to logarithms) and describes the physics of DIS scattering qualitatively, shows that the form factor data do not necessarily
demand that the nucleon wave function include  $L>0$ nonspherical angular momentum components (although they are certainly not ruled out).

In forthcoming work, including Ref.~\cite{RPG}, the structure of the nucleon wave function will be
used to describe the transitions to other baryons and to excited states of the nucleon. For that purpose a
modified version of the diquark propagator $D^{\mu \nu}$
will be used.  These studies will give more definitive information about angular momentum components in both the nucleon and other baryons.

\subsection{Shape of the nucleon} \label{sec:def}

We now turn to the interesting discussion of the shape of the nucleon.  Shapes of nuclei have been discussed for over 50 years, and are still an active area of research.   The Contemporary Physics Education Project (CPEP) \cite{CPEP} may be illustrative of the discussion found in the recent popular literature. In the discussion of their nuclear science wall chart, they refer to nuclear shape without giving a definition.   In an early paper, Rainwater \cite{Rainwater}  relates nuclear shape to the presence of a nuclear quadrupole moment, and this definition seems to be universal.  A very nice discussion of nuclear shapes and quadrupole moments can be found in the Nobel lectures by Rainwater \cite{Nrainwater}, Bohr \cite{Nbohr}, and Mottelson \cite{NMottelson}.

The quadrupole moment is determined by the charge density.  For example, consider a charged spin-1/2 quark moving about a fixed spin-1 particle (the diquark).  Since the diquark is merely a representation of the two quarks not being probed by the photon (we assume that there are no two-body charge operators), we can ignore its charge.    For definiteness, suppose the quark has angular momentum $\ell=1$ and the total angular momentum of the state is 1/2.  In nonrelativistic terms, the charge density of the spin-up state is then
\ba
\rho_e({\bf r})&=&e\,\psi^2(r)
\nonumber\\
&=&e\left[\sqrt{\sfrac23}\bra{-\sfrac12}Y^*_{11}-\sqrt{\sfrac13}\bra{\sfrac12}Y^*_{10}\right]\nonumber\\
&&\qquad\times\left[\sqrt{\sfrac23}\ket{-\sfrac12}Y_{11}-\sqrt{\sfrac13}\ket{\sfrac12}Y_{10}\right]
\nonumber\\
&=&\frac{e}{4\pi}\left[\sin^2\theta +\cos^2\theta\right]. \label{rhoNR}
\ea
This is a spherical result, even though the individual components that make up the wave function are not spherical.

In our model, the momentum space charge operator at $Q^2=0$ (both nucleons must be at rest to avoid Lorenz contraction effects) can be extracted from Eq.\ (\ref{eqF1})
\ba
\rho_e({\bf k})
               &=&3 \bar \Psi_N (P,k)
j_1\gamma_0 \Psi_N(P,k)
\nonumber\\
&=& \sfrac12 (1+\tau_3) |\psi (P,k)|^2,
\label{eqRhoE}
\ea
where $j_1$, defined in Eq.\ (\ref{eqmiccur}), is the quark charge operator.
Note that $\rho_e({\bf k})$ is proportional
to the nucleon charge operator and to
the square of the scalar wave function $\psi$
(a function of ${\bf k}^2$ only) as expected.
The electric charge density of the nucleon
is spherically symmetric.
The radial momentum distribution $|\psi(P,k)|^2$,
multiplied by the kinematic integration factor ${m_s^3}/(2E_s)$, 
is shown in Fig.\  \ref{fig_rho} for the two models
considered in this work.

The charge distribution derived in Ref.~I is also spherically symmetric, even though the wave functions are not.  This is because the angular dependence in the wave functions of Ref.~I (due to the angular dependence of the polarization vectors $\eta$) is canceled in much the same way as in example (\ref{rhoNR}).

We can also use the matter distribution to analyze
the nucleon shape.  In this case the operator $j_1$ is absent, and
\ba
\rho_m({\bf k}) &=& 3 \bar \Psi_N (P,k)\gamma_0 \Psi_N(P,k),
\nonumber \\
&=& 3 |\psi(P,k)|^2.
\label{eqRhoM}
\ea
This is also spherical.

If the shape is defined by the charge density, we conclude that the nucleon is spherical, unless it can be shown to exhibit some collective motion that would allow us to interpret it as a deformed state precessing about an axis different from its symmetry axis \cite{Rainwater, Nrainwater,Nbohr,NMottelson,Buchmann01}.  This would imply the existence of a rotational band and superlarge quadrupole radiation, which have not been observed.

Recently Miller \cite{Miller03} and Kvinikhidze and Miller \cite{Miller06}  introduced a spin direction dependent (SDD) density operator  as a means of describing the shape of the nucleon.  Nonrelativistically, the charge SDD  operator is
\ba
\rho^{{SDD}}_e({\bf r},{\bf n})=\rho_e({\bf r})\sfrac12[1+\sigma\cdot {\bf n}]
\ea
where ${\bf n}$ is the direction of the spin quantization.  Choosing ${\bf n}$ to be in the $+z$ direction, the spin-up matrix elements of this operator for the nonrelativistic case studied in Eq.\  (\ref{rhoNR}) are
\ba
\rho^{{SDD}}_e({\bf r},{\bf n})
&=&e\left[\sqrt{\sfrac23}\bra{-\sfrac12}Y^*_{11}-\sqrt{\sfrac13}\bra{\sfrac12}Y^*_{10}\right]
\sfrac12(1+\sigma_3)
\nonumber\\
&&\qquad\times\left[\sqrt{\sfrac23}\ket{-\sfrac12}Y_{11}-\sqrt{\sfrac13}\ket{\sfrac12}Y_{10}\right]
\nonumber\\
&=&\frac{e}{4\pi}\cos^2\theta, \label{rhosdd}
\ea
because, in this case, the spin projection operator has projected out the spin-up state of the quark, unveiling the angular momentum contained in $Y_{10}^2$.   Using this definition Kvinikhidze and Miller \cite{Miller06}  analyzed the relativistic nucleon model of Ref.~I, demonstrating that the SDD matrix elements generate angle-dependent terms in the nucleon SDD charge and matter densities.


Relativistically,  the SDD  density operator includes
the factor of $(\gamma_0 +{\bf n} \cdot \gammavec  \gamma_5)/2$,
where ${\bf n}$ is a unit vector that specifies the
quark spin direction.
With this definition, the SDD
electric charge distribution for a nucleon with polarization ${\bf \hat s}$ becomes
\ba
\rho_e^{SDD}({\bf k},{\bf n}, {\bf s})&=&
\frac{3}{2} \bar \Psi_N (P,k)  j_1
(\gamma_0 +{\bf n} \cdot \gammavec \gamma_5)
\Psi_N( P,k)
\nonumber  \\
&=&
\frac{1}{2}
\left[
\frac{1+\tau_3}{2}
+\frac{1+ 5\tau_3}{6}   {\bf n} \cdot {\bf \hat s}\right]
|\psi( P,k)|^2,  \nonumber \\
& &
\label{eqKMe0}
\ea
and for the matter distribution
\ba
\rho_m^{SDD}({\bf k},{\bf n}, {\bf s})&=&
\frac{3}{2} \bar \Psi_N ( P,k)
(\gamma_0 +{\bf n} \cdot \gammavec \gamma_5)
\Psi_N( P,k)
\nonumber  \\
& = &
\frac{1}{2}(3+{\bf n} \cdot {\bf \hat s}) |\psi( P,k)|^2
\label{eqKMm}
\ea
where, as before, $\psi(P,k)$ is independent of angles in the nucleon rest frame.
For the model presented in this paper, these distributions are also spherically symmetric, reflecting the fact that the model contains only $S$-wave components.

What are the larger implications of these observations?  The SSD operators are potentially quite interesting.  If they can be measured directly, they will reveal the angular momentum content of a state.  But using these operators to define the ``shape'' of a state  is contrary to what is usually understood as the shape (i.e., the charge or mass quadrupole moment density).  

\acknowledgements

It is a pleasure to thank John Arrington, who provided the reanalyzed $G_{Mp}$  data we used in our fit, and Christian Weiss, for helpful discussions of the quark distribution functions.  F.\ G.\ would also like to acknowledge helpful discussions at the Workshop on Nucleon Form factors (N05) held at the Laboratori Nazionali di Frascati, in October, 2005, and at the Workshop on Exclusive Reactions at High Momentum Transfer, held at Jefferson Lab in May, 2007.

This work began at the Jefferson Laboratory and was continued in Lisbon.  G.\ R.\ and M.\ T.\  P.\  want to thank Jefferson Laboratory and F.\ G.\ wants to thank the Centro de F\'\i sica Nuclear da Universidade 
de Lisboa, Portugal  for their hospitality. 
G.\ R.\ was supported by the portuguese Funda\c{c}\~ao para 
a Ci\^encia e Tecnologia (FCT) under the grant  
SFRH/BPD/26886/2006 and  POCTI/FNU/50358/2002.  This work was partly supported by Jefferson Science Associates, LLC under U.S. DOE Contract No. DE-AC05-06OR23177.

\appendix

\section{Evaluation of the hadronic tensor in the DIS limit}

Starting from the general formula (\ref{hadcurr}), the hadronic tensor requires evaluation of the following integral and trace
%
\bea
W^{\mu\nu}(q,P)=\sfrac32\sum_{s,s_1,s_2}\int\!\!\!\! \int_{p'k}
J^{\nu\dagger} J^\mu&&\cr
=\sfrac34\int\!\!\!\! \int_{p'k}
\psi^2(P,k)\Big\{j_1^2{\rm Tr}_0 &-&\sfrac13(\tau_ij_1^2\tau_i){\rm Tr}_1\Big\}\qquad
\eea
%
%
where the DIS current was defined in (\ref{DIScurrent}), the quark current $j_1$ in  (\ref{quarkff}), and the traces for contributions from the isospin 0 and 1 diquarks are (here we use the normalization $\bar u u=2M$)
\bea
{\rm Tr}_0&=&{\rm Tr}\bigg\{(M+\not\!P)\left(\gamma^\mu-\frac{\not\!q q^\mu}{q^2}\right)\not\!p' \left(\gamma^\nu-\frac{\not\!q q^\nu}{q^2}\right)\bigg\} \nonumber\\
{\rm Tr}_1&=&\sfrac13{\rm Tr}\bigg\{(M+\not\!P)\gamma^\alpha\gamma^5\left(\gamma^\mu-\frac{\not\!q q^\mu}{q^2}\right)\not\!p'
\nonumber\\
&&\qquad\qquad\times \left(\gamma^\nu-\frac{\not\!q q^\nu}{q^2}\right)\gamma^5\gamma^\beta\bigg\} D_{\alpha\beta}\, ,
\eea
and the phase-space integral was defined in (\ref{DISint}).  The quark mass has been dropped from the traces since its contribution is negligible in the DIS limit.  This tensor is covariant and may be calculated in any frame, and at this point we will not specify a frame.

The quark currents are
\bea
j_1^2=\lambda^2\left(\sfrac16+\sfrac12\tau_3\right)^2&=&\lambda^2\left(\sfrac5{18}+\sfrac16\tau_3\right)\nonumber\\
\sfrac13\tau_i\left(\sfrac5{18}+\sfrac16\tau_3\right)\tau_i&=&\sfrac5{18}-\sfrac1{18}\tau_3 \label{eq:quarkc}
\eea
The diquark polarization sum, defined for the case when $P_+=P_-+q$ in Eq.~(\ref{eqDeltaP2}), is now simply $-g_{\alpha\beta}+P_\alpha P_\beta/M^2$, from which we obtain immediately
\bea
{\rm Tr}_0=-{\rm Tr}_1&=&4\tilde P^\nu \tilde p'^\mu +4 \tilde p'^\nu \tilde P^\mu \nonumber\\
&&-4 (P\cdot p' )\left(g^{\mu\nu}-\frac{q^\nu q^\mu}{q^2}\right)\, ,
\eea
where, for any four-momentum,
\be
\tilde P^\mu\equiv \left(P^\mu -\frac{(P\cdot q)q^\mu}{q^2}\right)\, .
\ee
Hence the quark currents (\ref{eq:quarkc}) must be added, giving the current factor reported in Eq.~(\ref{eq:ei}).

To extract the structure functions $W_1$ and $W_2$, we go to a collinear frame (where ${\bf P}$ and ${\bf q}$ are in the $\hat z$ direction).   Then, since the wave function depends only on $(P-k)^2=(p'-q)^2$, and since the numerator is linear in $p'$, the  transverse components of $p'$ integrate to zero, and we can write
\bea
p'=AP+Bq
\eea
with $A$ given in Eq.~(\ref{eq:A}) and
\bea
B&=&\frac{(p'\cdot P)(P\cdot q)-M^2(p'\cdot q)}{(p\cdot q)^2+M^2Q^2}\, .
\eea
Since $\tilde q=0$, the trace terms give
\bea
{\rm Tr}_0=-{\rm Tr}_1&=& 8A\tilde P^\nu \tilde P^\mu - 4(P\cdot p')\left(g^{\mu\nu}-\frac{q^\nu q^\mu}{q^2}\right)\qquad
\eea
giving the expressions (\ref{eq:W12}) for the structure functions.

If  the integrals over $k_\pm$ are evaluated in the nucleon rest frame, as discussed in Sec.~\ref{sec:qdis}, the phase-space integral can be written  
\bea
\int\!\!\!\! \int_{p'k}
&=&\frac{x}{(1-x)Q^2}\int\frac{ d^2k_\perp}{2(2\pi)^2 } \int dk_+ dk_-\cr
&&\times\delta\left(k_+-\frac{m^2_s+k_\perp^2}{M(1-x)}
\right)\delta\big(k_--M(1-x)\big)\, ,\qquad
\eea
giving the limits 
\bea
P\cdot p' \to \frac{Q^2}{2x}\qquad A\to  x\, .
\eea
These lead directly to the expression (\ref{kpint}) for the quark distribution amplitude.

\section{Wave function normalization in  spectator and light-front theories}

To simplify the formulas in this section, we work with a wave function with only one pole:
\bea
\psi(P,k)\to \frac{N_0}{m_s(\beta+\chi-i\epsilon)}\, .
\eea
All results are also true if the wave function has two poles, as can be easily seen after the arguments have been developed.

Begin by examining a four-dimensional integral related to the normalization  integral.  At $Q^2=0$, the integral  is
\bea
I_4\equiv -i\int \frac{d^4 k}{(2\pi)^4} \frac{\psi^2(P_0,k)}{D_1}
= -i\int \frac{d^4 k}{(2\pi)^4} \frac{M^2 N_0^2}{D_1 D_2^2} \label{norm4d}
\eea
where the spectator four-momentum $k$ is unconstrained (so that the spectator is no longer on-shell) and the denominators therefore contribute two poles and two double poles in the complex $k_0$ plane:
\bea
D_1&=&m_s^2-k^2-i\epsilon=(E_s-k_0-i\epsilon)(E_s+k_0-i\epsilon)\nonumber\\
D_2&=&Mm_s\beta+|{\bf k}|^2+(M-m_s)^2 -(M-k_0)^2 -i\epsilon \nonumber\\
&=&(\sqrt{\xi + |{\bf k}|^2}+M-k_0-i\epsilon)\nonumber\\
&&\qquad\times(\sqrt{\xi + |{\bf k}|^2}-M+k_0-i\epsilon)\, ,\qquad
\eea
where $\xi=Mm_s\beta+(M-m_s)^2$.
The single poles are at $k_0=\pm E_s$ and the double poles are at $k_0=M\pm\sqrt{\xi+|{\bf k}|^2}$.  

The spectator theory organizes the infinite series of Feynman diagrams that describe (in this case) the quark-diquark interaction so that the spectator (diquark) is always on-shell; contributions from terms in which the spectator is off-shell are  included in higher order terms in the kernel.  In the present example this means that the {\it only\/} contribution from the integral (\ref{norm4d}) that we may include  comes from  the positive energy spectator pole, and other contributions to the four-dimensional integral would be included as part of the higher order terms in the kernel, and hence are part of the wave function $\psi$ that we are modeling.  Separating out the positive energy spectator pole from (\ref{norm4d}) gives the spectator normalization integral
\bea
I_s=\int \frac{d^3 k}{(2\pi)^32E_s} \psi^2(P_0,\hat k)\, ,\label{normspec}
\eea
where $\hat k^2=m_s^2$.

In light-cone theory the wave function is the solution of a generalized hamiltonian dynamics with $H_+=H+P_z$ the generalized hamiltonian.  The $+$ component of momentum is not conserved.  However, some of the features of the light-cone approach (and, in particular, the normalization issues discussed here) can be understood by evaluating the integral (\ref{norm4d}) using light-cone coordinates.  In these coordinates, the four-dimensional integral becomes
\bea
I_4
&=& -i\int \frac{dk_+\,dk_-\,d^2 k_\perp\;M^2 N_0^2}{2(2\pi)^4D_1'(D_2')^2} \label{norm4dlc}
\eea
where now
\bea
D'_1&=&m_s^2+k^2_\perp-k_-k_+-i\epsilon \nonumber\\
D_2'&=&\xi+k^2_\perp-(M-k_+)(M-k_-)-i\epsilon\, .
\eea
This integral has only one pole and one double pole in the complex $k_+$ plane.  If $0<k_-<M$, the single pole is in the lower half plane and the double pole is in the upper half plane, but outside of this region either the single or the double pole migrates so that all are in the same half plane, and the integral is zero \cite{Gross:1983yt,Chang:1968bh}.  Hence the exact answer is given by the spectator pole, with $0<k_-<M$.  Introducing $1-x=k_-/M$, the momentum fraction carried by the diquark, the integral becomes
\bea
I_4=\int_0^1\frac{dx}{1-x}\int\frac{d^2k_\perp}{2(2\pi)^3}\frac{M^2N_0^2}{x^2d^2}\label{normc}
\eea
with
\bea
d=\frac{\xi+k_\perp^2}{x}+\frac{m_s^2+k_\perp^2}{1-x} -M^2\, .
\eea
 
 At first glance, it might appear that the light-cone integral (\ref{normc}) has nothing to do with the spectator integral (\ref{normspec}), but an interesting connections was noticed a long time ago.  To make this connection explicit, we transform the spectator integral  into light-cone variables by transforming $k_z$ to $x$ using
\bea
M(1-x)&=&E_s-k_z\nonumber\\
Mdx&=&\left(1-\frac{k_z}{E_s}\right)dk_z =M(1-x)\frac{dk_z}{E_s}\nonumber\\
M(E_s+k_z)&=&(m_s^2+k_\perp^2)/(1-x)\nonumber\\
\hat D_2\equiv D_2(\hat k)&=& \xi+k_\perp^2 +k_z^2-(M-E_s)^2 \nonumber\\
&=&x\,d\, .
\eea
This gives the result
\bea
I_s=\int_{-\infty}^1\frac{dx}{1-x}\int\frac{d^2k_\perp}{2(2\pi)^3}\frac{M^2N_0^2}{x^2d^2}\, ,
\eea
as discussed in Sec.~\ref{sec:norm}.  

The difference in support of the light-cone and spectator integrals is due to the different way the singularities of the triangle diagram are handled.  An analysis of the advantages and disadvantages of these two methods is beyond the scope of this paper and will be discussed in a future work.  Here note only that from the standpoint of the spectator method, the integral over $k_z$ is limited by the light-cone requirement that $0<x$, which translates into the limit $0<E_s-k_z$, implying
\bea
\frac{m_s^2+k_\perp^2-M^2}{2M}<k_z\, .
\eea
This limit breaks the symmetry between the components of ${\bf k}$, with $k_z$ treated differently from the other two components.  If one is not careful, the light-cone method breaks rotational invariance, and can cause problems at low energy where manifest rotational invariance is an important constraint.

\section{Details of the construction of the two-pion cut term, Eq.~(\ref{dGcloud})}

Unfortunately, chiral calculations of the nucleon form factors \cite{Kaiser:2003qp,Hammer:2003qv} only give information about the imaginary part of the isovector form factors in the timelike region, over a small range of $Q^2$ near the onset of the two-pion cut (in the region of $4m_\pi^2<t=q^2=-Q^2\alt 10 m_\pi^2$).
The real parts of the form factors for both spacelike 
and timelike regions are obtained using dispersion 
theory, and are sensitive to the 
high momentum behavior  ($t=-Q^2 >> M^2$)
of the imaginary part.

Hence, to use the results of the chiral calculations we chose a mathematical function that reproduced the chiral calculations of the imaginary part at small $t$, but gave a real part with a strength that could be varied.  A convenient function with a simple behavior in the complex plane is
\bea
g_{i-}(t)=\frac{b_i^2}{b_i\left(1-\frac{t}{m_1^2}\right)\left(1-\frac{t}{m_2^2}\right)-ia_i\frac{\gamma(t)}{t}} 
\label{cloudterm}
\eea
with the width function 
\be
\gamma(t)=\frac{\mu^2q_{\pi\pi}^3}{\left(1-\frac{t}{2\mu^2}\right)}\, ,
\ee
where $\mu\equiv m_\pi$ and $q_{\pi\pi}=\sqrt{t/\mu^2-4}$ is the relative momentum of the two pions (in units of $\mu$) in the $\pi\pi$ rest system.
Note that the width goes as $q_{\pi \pi}^3$, as required by the $P$-wave nature of the $\rho$, and that $g_{i-}(0)=0$, so $g_{i-}$ does not contribute to the charge or magnetic moments.  The imaginary part of $g_{i-}$ for $t>4\mu^2$ is
\bea
{\rm Im}g_{i-}(t)&=&\frac{b_i^2a_i\gamma(t)/t}{b_i^2\left(1-\frac{t}{m_1^2}\right)^2\left(1-\frac{t}{m_2^2}\right)^2+a_i^2\left(\frac{\gamma(t)}{t}\right)^2} \nonumber\\
&\simeq& a_i\frac{\gamma(t)/t}{E(t)}\left\{1 - \frac{a_i^2}{b_i^2}\frac{\gamma^2(t)}{t^2E(t)}+\cdots\right\}
\eea
where
\be
E(t)\equiv\left(1-\frac{t}{m_1^2}\right)^2\left(1-\frac{t}{m_2^2}\right)^2\,. 
\ee
The expansion holds for low $q_{\pi \pi}$ momentum 
($q_{\pi \pi} << M$), when   $E(t) \simeq 1$. 
This shows that ${\rm Im}g_{i-}(t)$ is independent of $b_i$ to lowest order, as shown in in Figs.~\ref{fig:imE} and \ref{fig:imM}.

As it turns out, the simple function $\gamma(t)/t$ does an excellent job 
of fitting the chiral calculations, provided we choose $a_E=0.17$ and   $a_M=0.68$. 
The mass $m_1^2=28 \mu^2$ was chosen to equal the square of the $\rho$ mass, and $m_2^2=50 \mu^2$ is large enough to provide the needed high momentum convergence without seriously affecting the shape of the imaginary part in the low $t$ region.  

Finally,  (\ref{cloudterm}) may be analytically continued to $t=-Q^2<0$.  Giving $t$ a small positive imaginary part and using
\be
i\sqrt{(t+i\epsilon)/\mu^2-4}\to\sqrt{4+Q^2/\mu^2}
\ee
gives the result ({\ref{dGcloud}).



\end{document}